# Statistical Estimation of Malware Detection Metrics in the Absence of Ground Truth

Pang Du, Zheyuan Sun, Huashan Chen, Jin-Hee Cho, *Senior Member, IEEE*, and Shouhuai Xu

***Abstract*— The accurate measurement of security metrics is a critical research problem, because an improper or inaccurate measurement process can ruin the usefulness of the metrics. This is a highly challenging problem, particularly when the ground truth is unknown or noisy. In this paper, we measure five malware detection metrics in the *absence* of ground truth, which is a realistic setting that imposes many technical challenges. The ultimate goal is to develop principled, automated methods for measuring these metrics at the maximum accuracy possible. The problem naturally calls for investigations into statistical estimators by casting the measurement problem as a *statistical estimation* problem. We propose statistical estimators for these five malware detection metrics. By investigating the statistical properties of these estimators, we characterize when the estimators are accurate, and what adjustments can be made to improve them under what circumstances. We use synthetic data with known ground truth to validate these statistical estimators. Then, we employ these estimators to measure five metrics with respect to a large data set collected from VirusTotal.**

## I. Introduction

THE importance of security metrics has been well appreciated, despite the slow progress towards the ultimate goal. However, the measurement of security metrics is little understood. This fact may be attributed to the deceptive simplicity of the problem that one may get at a first glance. Under the premise that the ground truth is known (e.g., which files are malicious or benign), it is indeed straightforward to obtain the values of security metrics.

However, in practice, we often encounter situations in which the ground truth is either unknown or noisy. The lack of ground truth has been recognized as a tough hurdle that prevents security metrics from being measured accurately. For example, machine learning (or data mining) based cyber defense approaches often need to train defense models from datasets with known ground truth. In order to obtain the ground truth, two approaches have been widely used. The first approach relies on the manual examination by human experts. However, this approach is not scalable because limited numbers of human experts often need to examine a much larger number of objects. This approach is also error-prone due to the cognitive limitations in dealing with a large workload and/or inherent judgment errors. The second approach depends on third-party information, such as a list of blacklisted websites [1], [2]. However, the third-party information may be outdated, as evidenced by the practice that researchers would have to vet such information by some means [3], [4]. Indeed, this approach only defers the problem to the third party and thus does not resolve the fundamental issue.

As a consequence, it has become a popular practice to derive a *pretended* ground truth by conducting some kinds of voting by a set of sources (e.g., detectors). However, the validity and reliability of this rule-of-thumb practice is little understood. This is particularly true in the context of malware detection, where a set of files (or objects) are labeled by multiple malware detectors [5]–[7].

Kantchelian *et al.* [8] appear to be the first to investigate this problem. Specifically, they investigate both unsupervised and supervised learning approaches to the aggregation of the labels given by multiple malware detectors into a single one. The setting in their unsupervised learning approach is similar to the one considered in the present paper. They propose using the naïve Bayes method and treat the unknown ground truth labels as hidden variables, using the Expectation-Maximization (EM) approach [9], [10] to estimate the metrics (e.g., false positive probabilities), and using the estimated metrics to infer the hidden ground truth labels. Their work [8] makes the following assumptions: (i) a detector has the same false positive probability with respect to any benign file and the same false negative probability with respect to any malicious file; (ii) the detectors label samples independent of each other; (iii) the fraction of malicious samples is centered around 0.5; and (iv) the detectors have low false positive probabilities and high false negative probabilities. Note that assumptions (iii) and (iv) are imposed by their choice of prior distributions, which is a necessary step in any Bayesian statistical analysis (i.e., these two assumptions are inherent to the approach they adopt). In the present paper, we initiate the investigation of the problem with a *different* approach, which only makes two of those four assumptions, namely assumptions (i) and (ii).

Manuscript received September 10, 2017; revised February 18, 2018 and April 18, 2018; accepted April 19, 2018. Date of publication May 4, 2018; date of current version June 5, 2018. This work was supported in part by the U.S. National Science Foundation (NSF) under Grant DMS-1620945, in part by the U.S. Department of Defense (DoD) through the Office of the Assistant Secretary of Defense for Research and Engineering, and in part by ARO under Grant W911NF-17-1-0566. The views and opinions of the authors do not reflect those of the NSF or DoD. The associate editor coordinating the review of this manuscript and approving it for publication was Prof. David Starobinski. *(Corresponding author: Shouhuai Xu.)*

P. Du is with the Department of Statistics, Virginia Tech, Blacksburg, VA 24061 USA.

Z. Sun, H. Chen, and S. Xu are with the Department of Computer Science, University of Texas at San Antonio, San Antonio, TX 78249 USA (e-mail: shxu@cs.utsa.edu).

J.-H. Cho is with the U.S. Army Research Laboratory, Adelphi, MD 20783 USA.

Color versions of one or more of the figures in this paper are available online at http://ieeexplore.ieee.org.

### A. Key Contribution

We make the following contributions in this work:

1) We bring up the research problem of measuring security metrics in the *absence* of ground truth. We propose formulating the measurement of security metrics as a *statistical estimation* problem. As a first step towards ultimately tackling this problem, we investigate the measurement of malware detection metrics in the setting where each file is examined by multiple malware detection tools (or *detectors* for short), which may have different capabilities in detecting malware. This is the de facto practice introduced by VirusTotal [11].
In order to solve the statistical estimation problem with *heterogeneous* malware detectors, we propose a statistical methodology with three steps: (i) design *naïve estimators* to estimate security metrics in question; (ii) investigate the statistical properties of these estimators, including their asymptotic distributions and bias; and (iii) design adjusted estimators while characterizing their applications.
2) In order to demonstrate the usefulness of the methodology, we propose statistical estimators for measuring the following malware detection metrics: (i) the percentage of malicious files in a population; (ii) the false positive probability of a malware detector; (iii) the false negative probability of a malware detector; and (iv) the trustworthiness of a malware detector defined by a pair of metrics, including the Bayesian detection probability (i.e., the probability that a file is malicious when a detector detects it as malicious) and the negative predictive probability (i.e., the probability that a file is benign when a detector detects it as benign). Using a large, real dataset provided by VirusTotal, we measure these metrics corresponding to the simple method of *majority voting* by malware detectors.
The following findings are identified in this paper: (i) A defender should use as many detectors as possible as long as these detectors do more good than harm (i.e., their false positive and false negative probabilities are smaller than 0.5). (ii) A defender can use a few (e.g., 5) "very good" detectors (i.e., their false positive and false negative probabilities are smaller than 0.1). However, our analysis of the real dataset shows that *none* of the detectors, including popular ones, satisfies this condition. (iii) Given a relatively small number of detectors whose qualities (i.e., false positive and false negative probabilities) are not known, a defender should use adjusted estimators rather than naïve estimators.

This paper is structured as follows. Section II describes the statistical estimation problem. Section III presents statistical estimators for the realistic setting of heterogeneous detectors and validates these estimators using synthetic data with known ground truth. Section IV discusses how the methodology is applied to analyze a real-world dataset. Section V discusses several important issues, including real-world implications and limitation of the present study. Section VI reviews related work. Section VII concludes the paper and discusses open future research problems. For better readability, we defer proofs of theorems and some algorithm details to the Appendix. The code that was used to produce the experimental results is available at https://github.com/Chenutsa/trustworthiness.

*Notations:* We summarize the notations as follows:

- $I$: The indicator function
- Pr: The probability function
- $m$: The number of files (or objects) in an experiment
- $n$: The number of (malware) detectors in an experiment
- $\mathcal{I}_0$, $\mathcal{I}_1$: The index sets of the benign and malicious files among the $m$ files ($|\mathcal{I}_0| = m_0$, $|\mathcal{I}_1| = m_1$), respectively
- $\pi_1$: The fraction of malicious files ($\pi_1 = m_1/m$)
- $\tilde{}$, $\hat{}$: For a parameter (e.g., $\pi_1$), $\tilde{}$ and $\hat{}$ respectively represents its naïve and adjusted estimators (i.e., $\tilde{\pi}_1$ and $\hat{\pi}_1$)
- $a_i$: The unknown, *true label* (i.e., ground truth) of the $i$-th file (or file $i$): $a_i = 0$ means benign and $a_i = 1$ means malicious
- $X_{ij}$: The label assigned to the $i$-th file by the $j$-th detector (or detector $j$)
- $Y_i$: The *voted label* of the $i$-th file by the $n$ detectors
- $\mu_1$, $\sigma_1^2$: The asymptotic mean and variance of $\tilde{\pi}_1$
- $p_{+j}$, $p_{-j}$: The false positive and false negative probabilities of the $j$-th detector, respectively.
- $q_{+j}, q_{-j}$: The trustworthiness of the $j$-th detector; $q_{+j}$ is known as *precision* or *Bayesian detection probability*, while $q_{-j}$ is known as *negative predictive probability*.
- $p_{ab}$: The probability the *voted label* of a file is $b$ under the condition that the file's unknown, *true label* is $a$.
- $Z_{ij}^{(ab)}$: The indicator variable that the $i$-th file has a *voted label* $a$ and a label $b$ assigned by the $j$-th detector
- $p_{c,ab,j}$: The probability that a file has a *voted label* $a$ and a label $b$ assigned by the $j$-th detector under the condition that the file's unknown, *true label* is $c$
- $\mu_{c,ab,j}$, $\sigma^2_{c,ab,j}$: The mean and variance of the number of files, each of which has a *true label* $c$, a *voted label* $a$, and a label $b$ assigned by the $j$-th detector
- $\rho_{c,ab,a'b',j}$: The co-variance between (i) the random variable representing the number of files, each of which has a *true label* $c$, a *voted label* $a$, and a label $b$ assigned by the $j$-th detector, and (ii) the random variable representing the number of files, each of which has a *true label* $c$, a *voted label* $a'$, and a label $b'$ assigned by the $j$-th detector

## II. Problem Statement and a Generic Characterization

### A. Problem Statement

Suppose there are $n$ malware detectors, called *detectors* for short, and there are $m$ unlabeled files (or objects). In the context of malware detection, detectors label files as benign or malicious. Let $a_i$ be the unknown, *true label* (i.e., ground truth) of the $i$-th file, with $a_i = 0$ indicating a benign file and $a_i = 1$ indicating a malicious file. Let

$$\mathcal{I}_0 = \{1 \le i \le m : a_i = 0\} \quad \text{and} \quad \mathcal{I}_1 = \{1 \le i \le m : a_i = 1\}$$

denote the set of the indices of the ground-truth benign and malicious files, respectively. Let $m_0$ and $m_1$ denote the number of ground-truth benign and malicious files, respectively, where $m_0 = |\mathcal{I}_0|$, $m_1 = |\mathcal{I}_1|$, and $m_0 + m_1 = m$.

Denote by $X_{ij}$ the label assigned to the $i$-th file by the $j$-th detector, where $X_{ij} = 0$ indicates that the $i$-th file is detected by the $j$-th detector as benign and $X_{ij} = 1$ indicates that the $i$-th file is detected by the $j$-th detector as malicious. Denote by $Y_i$ the *voted label*, namely the label assigned to the $i$-th file through a voting method by the $n$ detectors.

In this paper, we focus on *majority* voting, namely that $Y_i$ is determined by

$$Y_i = I\left(\sum_{j=1}^{n} X_{ij} \geq \frac{n}{2}\right), \quad 1 \leq i \leq m, \tag{1}$$

where $I(\cdot)$ is the indicator function. Note that when $\sum_{j=1}^{n} X_{ij} = n/2$, we set $Y_i = 1$ rather than $Y_i = 0$. The treatment of setting $Y_i = 0$ instead is similar because of the symmetry. As shown in Section III, this choice has no significant side-effect because the estimators are accurate. Recall that $Y_i = 0$ means the $i$-th file is treated as benign and that $Y_i = 1$ means the $i$-th file is treated as malicious. That is, $Y_1, \ldots, Y_m$ are the voting results, which are not necessarily the same as the *ground truth* $a_1, \ldots, a_m$.

*Definition 1 (Malware detection metrics): We are interested in the following malware detection* metrics*:*

- $\pi_1$*: This is the unknown portion of malicious files among the m files, defined as* $\pi_1 = m_1/m = |\mathcal{I}_1|/m$.
- $p_{+j}$ *and* $p_{-j}$*: These are the* unknown *false positive probability* $p_{+j}$ *and the false negative probability* $p_{-j}$ *of the* $j$*-th* $(1 \leq j \leq n)$ *detector.*
- $q_{+j}$ *and* $q_{-j}$*: These refer to the* trustworthiness *of the* $j$*-th* $(1 \leq j \leq n)$ *detector.*

*Remark 1: Note that* $q_{+j}$ *is also called* positive predictive value *or* precision *or* Bayesian detection rate*, while* $q_{-j}$ *is also called* negative predictive value*. These two metrics capture the following: A defender cares about the trustworthiness of a decision made by a detector, namely the probability that a file is indeed malicious (or benign) when a detector says it is malicious (or benign), respectively. Metrics* $p_{+j}$*,* $p_{-j}$*,* $q_{+j}$*, and* $q_{-j}$ *are defined in a* probabilistic *fashion, rather than the popular* empirical *fashion [12], [13], for two reasons: (i) Probabilistic metrics reflect the intrinsic capabilities of the detectors, which are* invariant *of the datasets in question; whereas, empirical definitions are specific to datasets and may vary from one dataset to another; and (ii) statistical estimators make sense only with respect to fixed quantities (e.g., probabilities), and do not work for random empirical quantities.*

The *research problem* is to estimate the malware detection *metrics*, $\pi_1$, $p_{+j}$, $p_{-j}$, $q_{+j}$, and $q_{-j}$, given a set of $m$ files and $n$ detectors. To simplify the measurement of metrics $p_{+j}$, $p_{-j}$, $q_{+j}$, and $q_{-j}$, we make the following assumption:

*Assumption 1 (Mistake Probabilities of Individual Detectors): For a specific detector* $j$*,* $1 \leq j \leq n$*, we assume*

$$\Pr(X_{ij} = 1 | a_i = 0) \text{ is the same } \forall i \in \mathcal{I}_0, \tag{2}$$

$$\Pr(X_{ij} = 0 | a_i = 1) \text{ is the same } \forall i \in \mathcal{I}_1, \tag{3}$$

*Eq.* (2) *says that the* $j$*-th detector has the same false positive probability when classifying* any *benign file as malicious. Eq.* (3) *says that the* $j$*-th detector has the same false negative probability when classifying* any *malicious file as benign.*

Under Assumption 1 mentioned above, $p_{+j}$, $p_{-j}$, $q_{+j}$ and $q_{-j}$ are obtained by

$$\begin{aligned} p_{+j} &= \Pr(X_{ij} = 1 | a_i = 0), \quad \forall i \in \mathcal{I}_0, \\ p_{-j} &= \Pr(X_{ij} = 0 | a_i = 1), \quad \forall i \in \mathcal{I}_1, \\ q_{+j} &= \Pr(a_i = 1 | X_{ij} = 1), \\ q_{-j} &= \Pr(a_i = 0 | X_{ij} = 0). \end{aligned}$$

*Remark 2: Assumption 1 implies that benign files have the same probability to be mislabeled by a* specific *detector as malicious. Similarly, malicious files have the same probability to be mislabeled by a* specific *detector as benign. This may not be universally true because some detectors may have a greater capability in detecting some types of malware than detecting other types of malware. Therefore, this assumption needs to be further investigated in the future.*

## *B. A Generic Characterization*

For detectors $j = 1, \ldots, n$ with respective false positive probabilities $p_{+j}$ and false negative probabilities $p_{-j}$, let us look at the probability that the majority voting method correctly labels a file when the file is indeed malicious. Define this probability for any $i$, $1 \leq i \leq m$, as

$$p_{11} = \Pr(Y_i = 1 | a_i = 1) = \Pr\left(\sum_{j=1}^{n} X_{ij} \geq \frac{n}{2} \middle| a_i = 1\right), \tag{4}$$

where $p_{11}$ is independent of $i$ because, under Assumption 1, the $X_{ij}$'s do not depend on the specificity of a malicious file.

Let $p_{-max}$ be the maximum false negative probability among the $n$ detectors, namely $p_{-max} = \max\{p_{-j} : 1 \leq j \leq n\}$. Then, $p_{11}$ can be given by

$$p_{11} \geq \Pr(Y \geq \frac{n}{2}) = \sum_{k=\lceil n/2 \rceil}^{n} \binom{n}{k} (1 - p_{-max})^k p_{-max}^{n-k}, \tag{5}$$

where $Y$ is a Binomial$(n, p_{-max})$ random variable and $\lceil x \rceil$ is the ceiling function.

Similarly, let $p_{-min}$ be the minimum false negative probability among the $n$ detectors, namely $p_{-min} = \min\{p_{-j} : 1 \leq j \leq n\}$. Then, $p_{11}$ can be given by

$$p_{11} \leq \sum_{k=\lceil n/2 \rceil}^{n} \binom{n}{k} (1 - p_{-min})^k p_{-min}^{n-k}. \tag{6}$$

## *C. The Case of Homogeneous* $p_{+j}$*'s and Homogeneous* $p_{-j}$*'s*

In order to draw insights into the majority voting method, we consider the special case of *homogeneous* detectors with the same false negative probability $p_-$, namely $p_- = p_{-1} = \ldots = p_{-n}$. In this case, we can obtain $p_{11}$ by

$$p_{11} = \sum_{k=\lceil n/2 \rceil}^{n} \binom{n}{k} (1 - p_-)^k p_-^{n-k}. \tag{7}$$

Fig. 1 demonstrates the values of $p_{11}$ with varying $n$ and $p_-$. The preceding discussion indicates that these numbers provide the lower bound (i.e., $p_{-max}$) and the upper bound (i.e., $p_{-min}$) for the $p_{11}$'s of the majority voting method with *heterogeneous* detectors of different false negative probabilities. The findings are highlighted as the following insights, while noting that similar insights can be drawn regarding $p_{00}$.

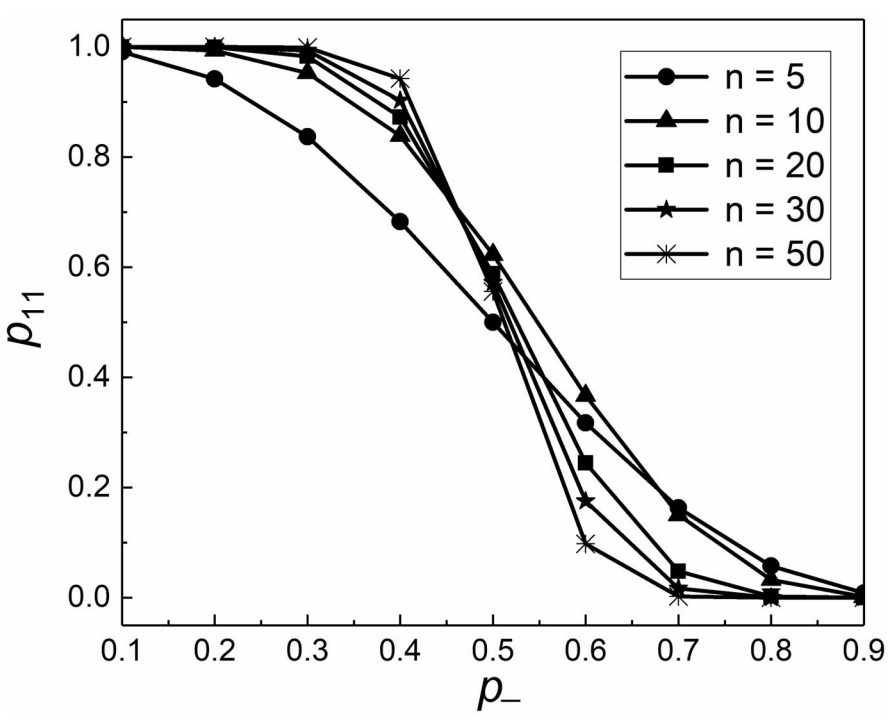

Fig. 1. Example $p_{11}$ of the majority voting method with $n$ homogeneous detectors of false negative probability $p_-$.

*Insight 1: In the special case of homogeneous $p_{-j}$'s,*

- *the majority voting method with $n \geq 20$ "good" detectors that have a small false negative probability $p_- < 0.2$ is almost perfect in detecting malware.*
- *the majority voting method with "fair" detectors ($0.2 \leq p_- < 0.5$) can still be almost perfect in detecting malware, as long as the number of such detectors is large enough (e.g., $n \geq 50$ for $p_- = 0.3$).*
- *the majority voting method with "poor" detectors ($p_- \geq 0.5$) is useless in detecting malware, no matter how many detectors are used.*

The preceding insights are equally applicable to the false positive probability because of the symmetry in the definitions.

## III. Estimators for Heterogeneous $p_{+j}$'s and $p_{-j}$'s

### A. Methodology

We propose the following methodology to cope with the more realistic setting of heterogeneous $p_{+j}$'s and $p_{-j}$'s.

- Step 1: Design naïve estimators of $\pi_1$, $p_{+j}$, $p_{-j}$, $q_{+j}$ and $q_{-j}$, respectively denoted by $\tilde{\pi}_1$, $\tilde{p}_{+j}$, $\tilde{p}_{-j}$, $\tilde{q}_{+j}$ and $\tilde{q}_{-j}$.
- Step 2: Investigate the statistical properties of these naïve estimators, especially their asymptotic distributions for characterizing their bias (i.e., the expected difference between the estimated value and the true value).
- Step 3: Design adjusted estimators to incorporate corrections to their bias.

Each step of the methodology is elaborated below.

### B. Designing Naïve Estimators

We define the *naïve* estimators of $\pi_1$, $p_{+j}$, $p_{-j}$, $q_{+j}$, and $q_{-j}$ as:

$$\tilde{\pi}_1 = \frac{\sum_{i=1}^m I(Y_i = 1)}{m} = \frac{\sum_{i=1}^m Y_i}{m}, \tag{8}$$

$$\tilde{p}_{+j} = \frac{\sum_{i=1}^m I(X_{ij} = 1 \text{ and } Y_i = 0)}{\sum_{i=1}^m I(Y_i = 0)}, \tag{9}$$

$$\tilde{p}_{-j} = \frac{\sum_{i=1}^m I(X_{ij} = 0 \text{ and } Y_i = 1)}{\sum_{i=1}^m I(Y_i = 1)}, \tag{10}$$

$$\tilde{q}_{+j} = \frac{\sum_{i=1}^m I(X_{ij} = 1 \text{ and } Y_i = 1)}{\sum_{i=1}^m I(X_{ij} = 1)}, \tag{11}$$

$$\tilde{q}_{-j} = \frac{\sum_{i=1}^m I(X_{ij} = 0 \text{ and } Y_i = 0)}{\sum_{i=1}^m I(X_{ij} = 0)}. \tag{12}$$

### C. Investigating Statistical Properties of Naïve Estimators

In order to examine the bias of the *naïve* estimators given by Eqs. (8)-(12), we need to investigate their asymptotic distributions. For this purpose, we make the following assumption:

*Assumption 2 (Independence of Decisions): We assume that the detectors independently make their decisions, namely*

$$\Pr(X_{ij} = c \text{ and } X_{i'j'=c'}) = \Pr(X_{ij} = c) \cdot \Pr(X_{i'j'=c'})$$

*for all $i, i' = 1, \ldots, m$, and $j, j' = 1, \ldots, n$ with $(i, j) \neq (i', j')$. Note that the independence applies when two different detectors are applied to the same file (i.e., $i = i'$ and $j \neq j'$) and when the same detector is applied to two different files (i.e., $i \neq i'$ and $j = j'$).*

*1) Statistical Properties of $\tilde{\pi}_1$:* Similar to the definition of $p_{11}$ in Eq. (4) under Assumption 1, we can define the probability that the voted label of a file is $b$ under the condition that its unknown *true label* is $a$ as

$$p_{ab} = \Pr(Y_i = b | a_i = a) \quad \text{for} \quad a, b \in \{0, 1\}.$$

If $a_i = 1$ (i.e., the ground truth is that the $i$-th file is malicious), then the $X_{ij}$'s are independent Bernoulli$(1 - p_{-j})$ random variables for $1 \leq j \leq n$. Hence the $Y_i$'s are independent and identically distributed (IID) Bernoulli random variables with the following probability parameter

$$p_{11} = \Pr\left(\sum_{j=1}^n X_{ij} \geq \frac{n}{2} \middle| a_i = 1\right).$$

Similarly, if $a_i = 0$, the $X_{ij}$'s are independent Bernoulli$(p_{+j})$ random variables and the $Y_i$'s are IID Bernoulli random variables with the following probability parameter

$$p_{01} = \Pr\left(\sum_{j=1}^n X_{ij} \geq \frac{n}{2} \middle| a_i = 0\right).$$

Now we present Theorem 1, with its proof deferred to Appendix A for better readability.

*Theorem 1: When $m_0 \to \infty$ and $m_1 \to \infty$, $\tilde{\pi}_1$ asymptotically follows the Normal distribution with the following mean and variance:*

$$\mu_1 = \pi_1 p_{11} + (1 - \pi_1) p_{01},$$
$$\sigma_1^2 = \frac{\pi_1 p_{11}(1 - p_{11}) + (1 - \pi_1) p_{01}(1 - p_{01})}{m}.$$

*Remark 3:* Theorem 1 *implies that the asymptotic bias of estimator $\tilde{\pi}_1$ is $\mu_1 - \pi_1 = \pi_1(p_{11} - 1) + (1 - \pi_1)p_{01}$. This bias is small in magnitude (i.e., an absolute value) when the majority voting method results in a high $p_{11}$ and a low $p_{01}$, but can be large when $p_{11}$ is small or $p_{01}$ is large.*

*2) Statistical Properties of $\tilde{p}_{+j}$, $\tilde{p}_{-j}$, $\tilde{q}_{+j}$, and $\tilde{q}_{-j}$:* Denote by $Z_{ij}^{(ab)}$ the indicator variable (i.e., a random variable) that the voted label of the $i$-th file is $a$ and the label assigned to the $i$-th file by the $j$-th detector is $b$, namely

$$Z_{ij}^{(ab)} = I(X_{ij} = b \quad \text{and} \quad Y_i = a) \quad \text{for} \quad a, b \in \{0, 1\}.$$

Intuitively, the $Z_{ij}^{(ab)}$'s partition the $m$ files into four groups: the files with $Z_{ij}^{(01)} = 1$, which are voted as benign but labeled as malicious by the $j$-th detector; the files with $Z_{ij}^{(00)} = 1$, which are voted as benign and also labeled as benign by the $j$-th detector; the files with $Z_{ij}^{(10)} = 1$, which are voted as malicious but labeled as benign by the $j$-th detector; and the files with $Z_{ij}^{(11)} = 1$, which are voted as malicious and also labeled as malicious by the $j$-th detector.

Under Assumption 2, we can obtain

$$\begin{aligned} Z_{ij}^{(01)} &= I(X_{ij} = 1 \quad \text{and} \quad Y_i = 0) \\ &= I(X_{ij} = 1) \cdot I\left(\sum_{k=1, k\neq j}^{n} X_{ik} < \frac{(n-2)}{2}\right), \\ Z_{ij}^{(00)} &= I(X_{ij} = 0 \quad \text{and} \quad Y_i = 0) \\ &= I(X_{ij} = 0) \cdot I\left(\sum_{k=1, k\neq j}^{n} X_{ik} < \frac{n}{2}\right), \\ Z_{ij}^{(10)} &= I(X_{ij} = 0 \quad \text{and} \quad Y_i = 1) \\ &= I(X_{ij} = 0) \cdot I\left(\sum_{k=1, k\neq j}^{n} X_{ik} \geq \frac{n}{2}\right), \\ Z_{ij}^{(11)} &= I(X_{ij} = 1 \quad \text{and} \quad Y_i = 1) \\ &= I(X_{ij} = 1) \cdot I\left(\sum_{k=1, k\neq j}^{n} X_{ik} \geq \frac{(n-2)}{2}\right). \end{aligned} \tag{13}$$

Note that for $a, b \in \{0, 1\}$, we have

$$\begin{aligned} &I(Y_i = a) \\ &\quad = I(X_{ij} = 1 \text{ and } Y_i = a) + I(X_{ij} = 0 \text{ and } Y_i = a) \\ &\quad = Z_{ij}^{(a1)} + Z_{ij}^{(a0)}, \text{ and} \\ &I(X_{ij} = b) \\ &\quad = I(X_{ij} = b \text{ and } Y_i = 1) + I(X_{ij} = b \text{ and } Y_i = 0) \\ &\quad = Z_{ij}^{(0b)} + Z_{ij}^{(1b)}. \end{aligned}$$

Hence Eqs. (9)-(12) can be respectively rewritten as

$$\tilde{p}_{+j} = \frac{\sum_{i=1}^{m} Z_{ij}^{(01)}}{\sum_{i=1}^{m} Z_{ij}^{(01)} + \sum_{i=1}^{m} Z_{ij}^{(00)}}, \tag{14}$$

$$\tilde{p}_{-j} = \frac{\sum_{i=1}^{m} Z_{ij}^{(10)}}{\sum_{i=1}^{m} Z_{ij}^{(10)} + \sum_{i=1}^{m} Z_{ij}^{(11)}}, \tag{15}$$

$$\tilde{q}_{+j} = \frac{\sum_{i=1}^{m} Z_{ij}^{(11)}}{\sum_{i=1}^{m} Z_{ij}^{(11)} + \sum_{i=1}^{m} Z_{ij}^{(01)}}, \tag{16}$$

$$\tilde{q}_{-j} = \frac{\sum_{i=1}^{m} Z_{ij}^{(00)}}{\sum_{i=1}^{m} Z_{ij}^{(00)} + \sum_{i=1}^{m} Z_{ij}^{(10)}}, \tag{17}$$

where $Z_{ij}^{(ab)}$ are defined in Eq. (13).

Denote by $p_{c,ab,j}$ the probability that a file has a voted label $a$ and a label $b$ assigned by the $j$-th detector under the condition that the file's unknown true label is $c$. That is,

$$p_{c,ab,j} = \Pr(Z_{ij}^{(ab)} = 1 | a_i = c) \quad \text{for} \quad a, b, c \in \{0, 1\}.$$

If $a_i = 1$, then the $X_{ij}$'s are independent Bernoulli$(1 - p_{-j})$ random variables for $1 \leq j \leq n$. According to Assumption 2, for each fixed triple $(a, b, j)$, the $Z_{ij}^{(ab)}$'s are IID Bernoulli random variables with the probability $p_{1,ab,j}$. The probabilities $p_{1,ab,j}$'s are obtained by

$$\begin{aligned} p_{1,01,j} &= \Pr(Z_{ij}^{(01)} = 1 | a_i = 1) \\ &= (1 - p_{-j}) \cdot \Pr\left(\sum_{k=1, k\neq j}^{n} X_{ik} < \frac{(n-2)}{2} \middle| a_i = 1\right), \\ p_{1,00,j} &= \Pr(Z_{ij}^{(00)} = 1 | a_i = 1) \\ &= p_{-j} \cdot \Pr\left(\sum_{k=1, k\neq j}^{n} X_{ik} < \frac{n}{2} \middle| a_i = 1\right), \\ p_{1,10,j} &= \Pr(Z_{ij}^{(10)} = 1 | a_i = 1) \\ &= p_{-j} \cdot \Pr\left(\sum_{k=1, k\neq j}^{n} X_{ik} \geq \frac{n}{2} \middle| a_i = 1\right), \\ p_{1,11,j} &= \Pr(Z_{ij}^{(11)} = 1 | a_i = 1) \\ &= (1 - p_{-j}) \cdot \Pr\left(\sum_{k=1, k\neq j}^{n} X_{ik} \geq \frac{(n-2)}{2} \middle| a_i = 1\right). \end{aligned}$$

Therefore, we obtain

$$\left(\sum_{i\in\mathcal{I}_1} Z_{ij}^{(01)}, \sum_{i\in\mathcal{I}_1} Z_{ij}^{(00)}, \sum_{i\in\mathcal{I}_1} Z_{ij}^{(10)}, \sum_{i\in\mathcal{I}_1} Z_{ij}^{(11)}\right),$$

which follows a multinomial distribution with parameters $m_1$ and $(p_{1,01,j}, p_{1,00,j}, p_{1,10,j}, p_{1,11,j})$. Actually, the components in this multinomial random vector respectively count the numbers of malicious files belonging to the aforementioned four groups determined by $Z_{ij}^{(ab)}$'s.

If $a_i = 0$, the $X_{ij}$'s are independent Bernoulli$(p_{+j})$ random variables for $1 \leq j \leq n$. Similarly, for each fixed triple $(a, b, j)$, the $Z_{ij}^{(ab)}$'s are IID Bernoulli random variables with

$$\begin{aligned} p_{0,01,j} &= \Pr(Z_{ij}^{(01)} = 1 | a_i = 0) \\ &= p_{+j} \cdot \Pr\left(\sum_{k=1, k\neq j}^{n} X_{ik} < \frac{(n-2)}{2} \middle| a_i = 0\right), \\ p_{0,00,j} &= \Pr(Z_{ij}^{(00)} = 1 | a_i = 0) \\ &= (1 - p_{+j}) \cdot \Pr\left(\sum_{k=1, k\neq j}^{n} X_{ik} < \frac{n}{2} \middle| a_i = 0\right), \\ p_{0,10,j} &= \Pr(Z_{ij}^{(10)} = 1 | a_i = 0) \\ &= (1 - p_{+j}) \cdot \Pr\left(\sum_{k=1, k\neq j}^{n} X_{ik} \geq \frac{n}{2} \middle| a_i = 0\right), \\ p_{0,11,j} &= \Pr(Z_{ij}^{(11)} = 1 | a_i = 0) \\ &= p_{+j} \cdot \Pr\left(\sum_{k=1, k\neq j}^{n} X_{ik} \geq \frac{(n-2)}{2} \middle| a_i = 0\right). \end{aligned}$$

Accordingly, we have

$$\left(\sum_{i\in\mathcal{I}_0} Z_{ij}^{(01)}, \sum_{i\in\mathcal{I}_0} Z_{ij}^{(00)}, \sum_{i\in\mathcal{I}_0} Z_{ij}^{(10)}, \sum_{i\in\mathcal{I}_0} Z_{ij}^{(11)}\right),$$

which follows a multinomial distribution with parameters $m_0$ and $(p_{0,01,j}, p_{0,00,j}, p_{0,10,j}, p_{0,11,j})$. Similarly, the components in this random vector respectively count the numbers of benign files belonging to the four groups determined by $Z_{ij}^{(ab)}$'s.

From Assumption 2, we know that the random vectors

$$\left(\sum_{i\in\mathcal{I}_1} Z_{ij}^{(01)}, \sum_{i\in\mathcal{I}_1} Z_{ij}^{(00)}, \sum_{i\in\mathcal{I}_1} Z_{ij}^{(10)}, \sum_{i\in\mathcal{I}_1} Z_{ij}^{(11)}\right) \quad \text{and}$$
$$\left(\sum_{i\in\mathcal{I}_0} Z_{ij}^{(01)}, \sum_{i\in\mathcal{I}_0} Z_{ij}^{(00)}, \sum_{i\in\mathcal{I}_0} Z_{ij}^{(10)}, \sum_{i\in\mathcal{I}_0} Z_{ij}^{(11)}\right)$$

are independent random variables. Hence, we have the following Theorem 2, with its proof deferred to Appendix A.

*Theorem 2: For $a, b, a', b', c \in \{0, 1\}$, we define*

$$\begin{aligned}\mu_{c,ab,j} &= m_c p_{c,ab,j},\\ \sigma^2_{c,ab,j} &= m_c p_{c,ab,j}(1 - p_{c,ab,j}), \quad \text{and}\\ \rho_{c,ab,a'b',j} &= -m_c p_{c,ab,j} p_{c,a'b',j}.\end{aligned}$$

*When $m_0 \to \infty$ and $m_1 \to \infty$, the following can be obtained:*

1) *The false positive probability estimator $\tilde{p}_{+j}$ asymptotically follows the Normal distribution with mean and variance being respectively*

$$\begin{aligned}\mu_{+j} &= \frac{\mu_{1,01,j} + \mu_{0,01,j}}{\mu_{1,01,j} + \mu_{1,00,j} + \mu_{0,01,j} + \mu_{0,00,j}},\\ \sigma^2_{+j} &= d_1^2\sigma^2_{1,01,j} + 2d_1d_2\rho_{1,01,00,j} + d_2^2\sigma^2_{1,00,j}\\ &\quad + d_1^2\sigma^2_{0,01,j} + 2d_1d_2\rho_{0,01,00,j} + d_2^2\sigma^2_{0,00,j},\end{aligned}$$

*where*

$$\begin{aligned}d_1 &= \frac{(\mu_{1,00,j} + \mu_{0,00,j})}{(\mu_{1,01,j} + \mu_{1,00,j} + \mu_{0,01,j} + \mu_{0,00,j})^2},\\ d_2 &= -\frac{(\mu_{1,01,j} + \mu_{0,01,j})}{(\mu_{1,01,j} + \mu_{1,00,j} + \mu_{0,01,j} + \mu_{0,00,j})^2}.\end{aligned}$$

2) *The false negative probability estimator $\tilde{p}_{-j}$ asymptotically follows the normal distribution with mean and variance being respectively*

$$\begin{aligned}\mu_{-j} &= \frac{\mu_{1,10,j} + \mu_{0,10,j}}{\mu_{1,10,j} + \mu_{1,11,j} + \mu_{0,10,j} + \mu_{0,11,j}},\\ \sigma^2_{-j} &= d_3^2\sigma^2_{1,10,j} + 2d_3d_4\rho_{1,10,11,j} + d_4^2\sigma^2_{1,11,j}\\ &\quad + d_3^2\sigma^2_{0,10,j} + 2d_3d_4\rho_{0,10,11,j} + d_4^2\sigma^2_{0,11,j},\end{aligned}$$

*where*

$$\begin{aligned}d_3 &= \frac{\mu_{1,11,j} + \mu_{0,11,j}}{(\mu_{1,10,j} + \mu_{1,11,j} + \mu_{0,10,j} + \mu_{0,11,j})^2},\\ d_4 &= -\frac{\mu_{1,10,j} + \mu_{0,10,j}}{(\mu_{1,10,j} + \mu_{1,11,j} + \mu_{0,10,j} + \mu_{0,11,j})^2}.\end{aligned}$$

3) *The positive predictive value estimator $\tilde{q}_{+j}$ asymptotically follows the Normal distribution with the following mean and variance*

$$\begin{aligned}\upsilon_{+j} &= \frac{\mu_{1,11,j} + \mu_{0,11,j}}{\mu_{1,01,j} + \mu_{1,11,j} + \mu_{0,01,j} + \mu_{0,11,j}},\\ \delta^2_{+j} &= e_1^2\sigma^2_{1,01,j} + 2e_1e_2\rho_{1,01,11,j} + e_2^2\sigma^2_{1,11,j}\\ &\quad + e_1^2\sigma^2_{0,01,j} + 2e_1e_2\rho_{0,01,11,j} + e_2^2\sigma^2_{0,11,j},\end{aligned}$$

*where*

$$\begin{aligned}e_1 &= \frac{\mu_{1,01,j} + \mu_{0,01,j}}{(\mu_{1,01,j} + \mu_{1,11,j} + \mu_{0,01,j} + \mu_{0,11,j})^2},\\ e_2 &= -\frac{\mu_{1,11,j} + \mu_{0,11,j}}{(\mu_{1,01,j} + \mu_{1,11,j} + \mu_{0,01,j} + \mu_{0,11,j})^2}.\end{aligned}$$

4) *The negative predictive value estimator $\tilde{q}_{-j}$ asymptotically follows the Normal distribution with the following mean and variance, respectively*

$$\begin{aligned}\upsilon_{-j} &= \frac{\mu_{1,00,j} + \mu_{0,00,j}}{\mu_{1,00,j} + \mu_{1,10,j} + \mu_{0,00,j} + \mu_{0,10,j}},\\ \delta^2_{-j} &= e_3^2\sigma^2_{1,00,j} + 2e_3e_4\rho_{1,00,10,j} + e_4^2\sigma^2_{1,10,j}\\ &\quad + e_3^2\sigma^2_{0,00,j} + 2e_3e_4\rho_{0,00,10,j} + e_4^2\sigma^2_{0,10,j},\end{aligned}$$

*where*

$$\begin{aligned}e_3 &= \frac{\mu_{1,10,j} + \mu_{0,10,j}}{(\mu_{1,00,j} + \mu_{1,10,j} + \mu_{0,00,j} + \mu_{0,10,j})^2},\\ e_4 &= -\frac{\mu_{1,00,j} + \mu_{0,00,j}}{(\mu_{1,00,j} + \mu_{1,10,j} + \mu_{0,00,j} + \mu_{0,10,j})^2}.\end{aligned}$$

*Remark 4: None of the means of the asymptotic distributions in Theorem 2 matches the true value of the corresponding metric, meaning that these naïve estimators are* biased *and their biases have more complicated forms than that of $\tilde{\pi}_1$.*

### D. Designing Adjusted Estimators

*1) Adjusting $\tilde{\pi}_1$ to $\hat{\pi}_1$:* From Theorem 1, we know that the asymptotic mean of $\tilde{\pi}_1$ is $\mu_1 = \pi_1 p_{11} + (1 - \pi_1)p_{01}$. By the *method of moments* [14], we can set up the following equation to obtain a new estimator of $\pi_1$:

$$\tilde{\pi}_1 = \pi_1 p_{11} + (1 - \pi_1)p_{01}. \tag{18}$$

By replacing $p_{11}$ and $p_{01}$ with their respective estimators $\tilde{p}_{11}$ and $\tilde{p}_{01}$, which can be obtained by the Monte Carlo method described in Section III-D4 with $p_{+j} = \tilde{p}_{+j}$ and $p_{-j} = \tilde{p}_{-j}$ for $j = 1, \ldots, n$, we can solve Eq. (18) to obtain an adjusted estimator of $\hat{\pi}_1$ as follows:

$$\hat{\pi}_1 = \frac{\tilde{\pi}_1 - \tilde{p}_{01}}{\tilde{p}_{11} - \tilde{p}_{01}}, \tag{19}$$

which has the bias removed. The adjusted estimator $\hat{\pi}_1$ is then used to compute

$$\hat{m}_1 = [m\hat{\pi}_1] \quad \text{and} \quad \hat{m}_0 = m - \hat{m}_1,$$

where $[x]$ is the rounding function (i.e., it returns a rounded integer of $x$). The estimators $\hat{\pi}_1$, $\hat{m}_1$ and $\hat{m}_0$ will be used in the estimator adjustments of $p_{+j}$, $p_{-j}$, $q_{+j}$, and $q_{-j}$.

*2) Adjusting $\tilde{p}_{+j}$ and $\tilde{p}_{-j}$ Respectively to $\hat{p}_{+j}$ and $\hat{p}_{-j}$:* The adjustments of $\tilde{p}_{+j}$ and $\tilde{p}_{-j}$ follow the same principle as, but are more complicated than, the adjustment of $\tilde{\pi}_1$ because $\tilde{p}_{+j}$ and $\tilde{p}_{-j}$ are related to all of the other estimators. We adjust one estimator $j$ at a time.

For adjusting $\tilde{p}_{+j}$ and $\tilde{p}_{-j}$, we define the following conditional probabilities that are conditioned on $a_i = 1$ and $a_i = 0$:

$$\begin{aligned}\alpha_{1j} &= \Pr\left(\sum_{k=1,k\neq j}^{n} X_{ik} < \frac{(n-2)}{2}\middle| a_i = 1\right),\\ \alpha_{2j} &= \Pr\left(\sum_{k=1,k\neq j}^{n} X_{ik} < \frac{n}{2}\middle| a_i = 1\right),\\ \beta_{1j} &= \Pr\left(\sum_{k=1,k\neq j}^{n} X_{ik} < \frac{(n-2)}{2}\middle| a_i = 0\right),\end{aligned}$$

$$\beta_{2j} = \Pr\left(\sum_{k=1,k\neq j}^{n} X_{ik} < \frac{n}{2}\bigg| a_i = 0\right).$$

For a fixed detector $j$, these conditional probabilities are in regard to the other $(n-1)$ detectors. Under the condition that $a_i = 1$ (i.e., the ground truth is that file $i$ is malicious), $\alpha_{1j}$ is the probability that no more than $(n-2)/2$ detectors other than $j$ detect file $i$ as malicious; $\alpha_{2j}$ is the probability that no more than $n/2$ detectors other than $j$ detect file $i$ as malicious. Correspondingly, $\beta_{1j}$ and $\beta_{2j}$ are condition probabilities under the condition that $a_i = 0$. We stress that these conditional probabilities are independent of $i$ because, under Assumption 1, these probabilities are the same for *any* $i$. These conditional probabilities can be estimated by their Monte Carlo estimators $\tilde{\alpha}_{hj}$ and $\tilde{\beta}_{hj}$, where $h = 1, 2$ with $p_{+k} = \tilde{p}_{+k}$ and $p_{-k} = \tilde{p}_{-k}$ for $k \neq j$. By Theorem 2, we can set up the following

$$\tilde{p}_{+j} = \frac{\hat{m}_1(1-p_{-j})\tilde{\alpha}_{1j} + \hat{m}_0 p_{+j}\tilde{\beta}_{1j}}{\hat{m}_1(1-p_{-j})\tilde{\alpha}_{1j} + \hat{m}_1 p_{-j}\tilde{\alpha}_{2j} + \hat{m}_0 p_{+j}\tilde{\beta}_{1j} + \hat{m}_0(1-p_{+j})\tilde{\beta}_{2j}},$$

$$\tilde{p}_{-j} = \frac{\hat{m}_1 p_{-j}(1-\tilde{\alpha}_{2j}) + \hat{m}_0(1-p_{+j})(1-\tilde{\beta}_{2j})}{\begin{pmatrix}\hat{m}_1 p_{-j}(1-\tilde{\alpha}_{2j}) + \hat{m}_1(1-p_{-j})(1-\tilde{\alpha}_{1j}) \\ +\hat{m}_0(1-p_{+j})(1-\tilde{\beta}_{2j}) + \hat{m}_0 p_{+j}(1-\tilde{\beta}_{1j})\end{pmatrix}}.$$

Some algebraic manipulation reduces these equations to the following linear equations of $p_{+j}$ and $p_{-j}$:

$$\begin{pmatrix} a_{11} & -a_{12} \\ a_{21} & -a_{22} \end{pmatrix}\begin{pmatrix} p_{+j} \\ p_{-j} \end{pmatrix} = \begin{pmatrix} b_1 \\ b_2 \end{pmatrix} \tag{20}$$

where

$$\begin{aligned}
a_{11} &= \hat{m}_0\tilde{\beta}_{1j}(1-\tilde{p}_{+j}) + \hat{m}_0\tilde{\beta}_{2j}\tilde{p}_{+j},\\
a_{12} &= \hat{m}_1\tilde{\alpha}_{1j}(1-\tilde{p}_{+j}) + \hat{m}_1\tilde{\alpha}_{2j}\tilde{p}_{+j},\\
a_{21} &= \hat{m}_0(1-\tilde{\beta}_{2j})(1-\tilde{p}_{-j}) + \hat{m}_0(1-\tilde{\beta}_{1j})\tilde{p}_{-j},\\
a_{22} &= \hat{m}_1(1-\tilde{\alpha}_{2j})(1-\tilde{p}_{-j}) + \hat{m}_1(1-\tilde{\alpha}_{1j})\tilde{p}_{-j},\\
b_1 &= \hat{m}_0\tilde{\beta}_{2j}\tilde{p}_{+j} - \hat{m}_1\tilde{\alpha}_{1j}(1-\tilde{p}_{+j}),\\
b_2 &= \hat{m}_0(1-\tilde{\beta}_{2j})(1-\tilde{p}_{-j}) - \hat{m}_1(1-\tilde{\alpha}_{1j})\tilde{p}_{-j}.
\end{aligned}$$

The adjusted estimators for $p_{+j}$ and $p_{-j}$ are then the solution $(\hat{p}_{+j}, \hat{p}_{-j})$ to Eq. (20).

*3) Adjusting $\tilde{q}_{+j}$ and $\tilde{q}_{-j}$ Respectively to $\hat{q}_{+j}$ and $\hat{q}_{-j}$:* Note that the positive and negative predictive values $q_{+j}$ and $q_{-j}$ are related to the false positive and negative probabilities $p_{+j}$ and $p_{-j}$ through $\pi_1$ as

$$q_{+j} = \frac{\pi_1(1-p_{-j})}{\pi_1(1-p_{-j}) + (1-\pi_1)p_{+j}}, \tag{21}$$

$$q_{-j} = \frac{(1-\pi_1)(1-p_{+j})}{(1-\pi_1)(1-p_{+j}) + \pi_1 p_{-j}}. \tag{22}$$

Given the adjusted estimators $\hat{\pi}_1$, $\hat{p}_{+j}$ and $\hat{p}_{-j}$, we can compute the adjusted estimators for $q_{+j}$ and $q_{-j}$ as

$$\hat{q}_{+j} = \frac{\hat{\pi}_1(1-\hat{p}_{-j})}{\hat{\pi}_1(1-\hat{p}_{-j}) + (1-\hat{\pi}_1)\hat{p}_{+j}}, \tag{23}$$

$$\hat{q}_{-j} = \frac{(1-\hat{\pi}_1)(1-\hat{p}_{+j})}{(1-\hat{\pi}_1)(1-\hat{p}_{+j}) + \hat{\pi}_1\hat{p}_{-j}}. \tag{24}$$

**Algorithm 1** Monte Carlo Computation of $\Pr\left(\sum_{j=1}^{n} X_{ij} \leq M\right)$ for Some Integer $M$

Input: A sequence of probabilities $\{p_j, j = 1, \ldots, n\}$, $N = 10,000$
Output: $\Pr\left(\sum_{j=1}^{n} X_{ij} \leq M\right)$ where $X_{ij} \sim \text{Bernoulli}(p_j)$
set a counter $k \leftarrow 0$.
**for** $\ell = 1 \text{to} N$ **do**
  generate $(x_{\ell 1}, \ldots, x_{\ell n})$ such that $x_{\ell j} \sim \text{Bernoulli}(p_j)$.
  **if** $\sum_{j=1}^{n} x_{\ell j} \leq M$ **then**
    $k \leftarrow k + 1$.
  **end if**
**end for**
Output $k/N$ as the final result.

*4) Monte Carlo Computation of $\Pr(\sum_{j=1}^{n} X_{ij} \leq M)$:* Note that $\Pr\left(\sum_{j=1}^{n} X_{ij} \leq M\right)$ is the probability that the total number of detectors that say that the $i$-th file is malicious is no greater than $M$, where $M$ is a given parameter. The calculation of the adjusted estimators requires one to estimate probabilities involving the sum of some of the $X_{ij}$'s, such as the probability estimators $\tilde{p}_{11}$, $\tilde{p}_{10}$, $\alpha_{1j}$, $\alpha_{2j}$, $\beta_{1j}$, and $\beta_{2j}$. Although the $X_{ij}$'s are independent Bernoulli random variables, the parameters of their distributions are different because the detectors' false positive probabilities and/or false negative probabilities can be different from each other (e.g., $p_{+j} \neq p_{+j'}$ where $j \neq j'$). This means that the sum of these variables does *not* follow any standard distribution. In this section, we describe a Monte Carlo method for evaluating probabilities involving such a form of variables. Without loss of generality, we present an algorithm for computing $\Pr\left(\sum_{j=1}^{n} X_{ij} \leq M\right)$ for any $M$. The computation of the other probabilities involving the sum of some $X_{ij}$'s is similar.

### E. Numerical Experiments

Here we present numerical experiments by using synthetic data with known ground truth to examine the accuracy of the estimators mentioned above. We consider two settings of false positive probabilities $p_{+j}$'s and false negative probabilities $p_{-j}$'s: *slight heterogeneity* and *true heterogeneity*.

*1) Experiments With Slightly Heterogeneous $p_{+j}$'s and $p_{-j}$'s:* In the first set of numerical experiments, we consider $n = 5, 10, 15, 25, 35$ detectors, whose $p_{+j}$'s and $p_{-j}$'s are *slightly heterogeneous*, by choosing them uniformly from the range $[\epsilon, \epsilon + 0.1]$, where $\epsilon = 0, 0.1, 0.2, \ldots, 0.7$. We fix $m = 50,000$ and $\pi_1 = 0.2$ in this set of numerical experiments.

For each fixed pair of parameters $(n, \epsilon)$, we generate $1,000$ samples as follows (see Algorithm 2 in Appendix B for details). We first randomly generate a set of $p_{+j}$'s and $p_{-j}$'s, each of size $n$, from the Uniform$(\epsilon, \epsilon + 0.1)$ distribution. We then generate 1,000 data samples for each pair $(n, \epsilon)$. To make the results comparable across the samples, we use the same $p_{+j}$'s and $p_{-j}$'s to generate the 1,000 data samples. Within each sample, we randomly generate the ground truth labels for the 50,000 files with exactly $50,000 \times 0.2 = 10,000$ malicious files and $50,000 - 10,000 = 40,000$ benign files. Then, the labels are assigned by the detectors and the voted labels are generated as in Section III-E1.

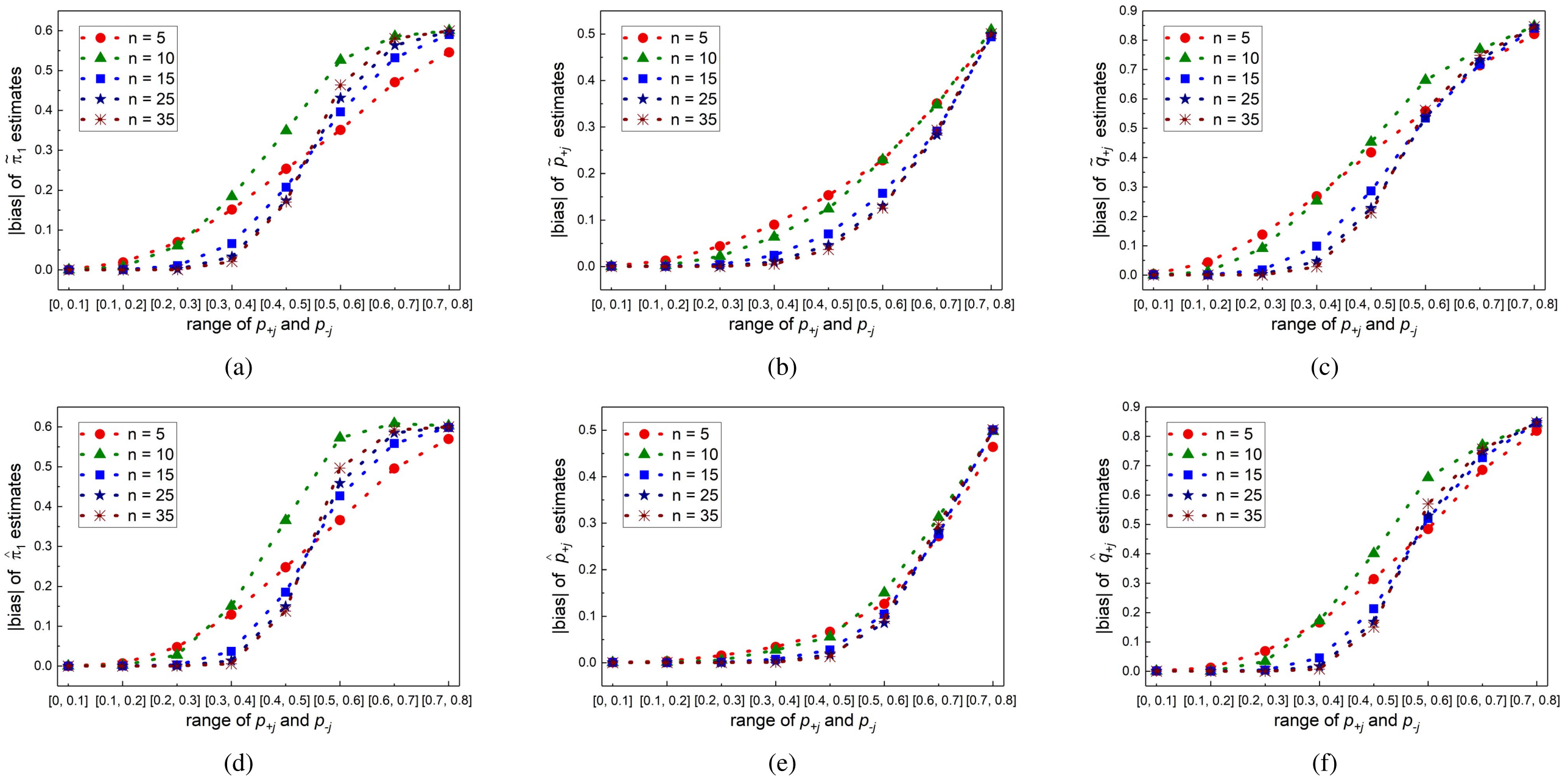

Fig. 2. Average biases (over the 1,000 data samples) of the estimators of $\pi_1, p_{+j}, q_{+j}$ in the numerical experiments with slightly heterogeneous $p_{+j}$'s and $p_{-j}$'s respectively chosen from the ranges as indicated (Section III-E1), where $j \in [1, n]$ and |bias| represents the absolute value of bias. (a) |bias| of $\tilde{\pi}_1$. (b) |bias| of $\tilde{p}_{+j}$. (c) |bias| of $\tilde{q}_{+j}$. (d) |bias| of $\hat{\pi}_1$. (e) |bias| of $\hat{p}_{+j}$. (f) |bias| of $\hat{q}_{+j}$.

Fig. 2 plots the biases of the naïve and adjusted estimators respectively for $\pi_1$, $p_{+j}$ and $q_{+j}$, and shows that the naïve and adjusted estimations are approximately the same. The results of the estimators for $p_{-j}$ and $q_{-j}$ are similar and thus omitted. We draw the following insights.

*Insight 2: In the case of* slightly heterogeneous *$p_{+j}$'s and $p_{-j}$'s,*

- *a small number (e.g., $n = 5$) of "very good" detectors (i.e., $p_{+j} < 0.1$ and $p_{-j} < 0.1$) can make the biases of naïve and adjusted estimators close to zero, meaning that majority voting is almost perfect in detecting malware.*
- *a large number (e.g., $n = 15$) of "good" detectors (i.e., $0.1 \leq p_{+j} < 0.2$ and $0.1 \leq p_{-j} < 0.2$) can make the biases of both the naïve and adjusted estimators close to zero, meaning that majority voting is almost perfect in detecting malware.*
- *a larger number (e.g., $n = 35$) of "fair" detectors (i.e., $0.2 \leq p_{+j} \leq 0.5$ and $0.2 \leq p_{-j} \leq 0.5$) make the biases of both the naïve and adjusted estimators close to zero, meaning that majority voting is almost perfect in detecting malware. Moreover, the adjusted estimators have lower biases, or are more accurate, than their corresponding naïve estimators.*
- *when the detectors are "poor" (i.e., $p_{+j} \geq 0.5$ and $p_{-j} \geq 0.5$), neither the naïve estimators nor the adjusted estimators are reliable because their biases are high.*

*2) Experiments With Truly Heterogeneous $p_{+j}$'s and $p_{-j}$'s:* In the second set of numerical experiments, we consider $m = 100,000$ files with a ground-truth portion of malicious files $\pi_1 = 0.58579$, which is chosen according to the $\pi_1$ derived from a preliminary analysis of the real dataset. We also generate synthetic data according to the adjusted estimators derived from the preliminary analysis. The preliminary analysis is not reported here, but its results are very similar to the analysis reported in Section IV. The intent is to make the synthetic data mimic the real data, but for the synthetic data we know the ground truth. Since the real dataset contains 47 detectors, we consider $n = 47$ simulated detectors. To accommodate heterogeneity, we choose $p_{+j}$ from the range of $(0.000617, 0.256)$ and $p_{-j}$ from the range of $(0.00238, 0.998)$, where $j \in [1, 47]$. To see the impact of "poor" detectors whose $p_{-j}$'s are greater than or equal to 0.5, we consider 8 "poor" detectors whose $p_{-j}$'s are respectively 0.515, 0.606, 0.648, 0.718, 0.732, 0.828, 0.921, and 0.998. We rank the $n = 47$ simulated detectors according to their $p_{-j}$'s in the decreasing order, which leads to a *fixed* list of simulated detectors named $1, 2, \ldots, 47$.

In order to see the impact of the number of detectors that are used in the majority voting method, we consider a sequence of experiments that respectively use the *first* $\kappa = 5, 15, 25, 35, 47$ detectors on the list of the 47 simulated detectors. This means, for example, that simulated detector 43 will be encountered only when we consider $\kappa = n = 47$ detectors, but detector 33 will be encountered when we consider $\kappa = 35 < n = 47$ and $\kappa = n = 47$ detectors. For a fixed $\kappa \in \{5, 15, 25, 35, 47\}$, we use these $\kappa$ detectors' $p_{+j}$'s and $p_{-j}$'s to generate $N = 1,000$ data samples. Within each sample, we randomly generate the ground truth labels for $m = 100,000$ files with exactly $100,000 \times 0.58579 = 58,579$ malicious files and $100,000 - 58,579 = 41,421$ benign files. Then each detector is applied to these $m$ files. When the true label of a file is 1 (malicious), the label assigned by the $j$-th detector is randomly generated according to the Bernoulli $(1 - p_{-j})$ distribution. When the true label of a file is 0 (benign), the label assigned by the $j$-th detector is randomly generated according to the Bernoulli ($p_{+j}$) distribution. Once all the $n$ detectors are applied to all the files, the voted label for a file is determined according to the majority voting method.

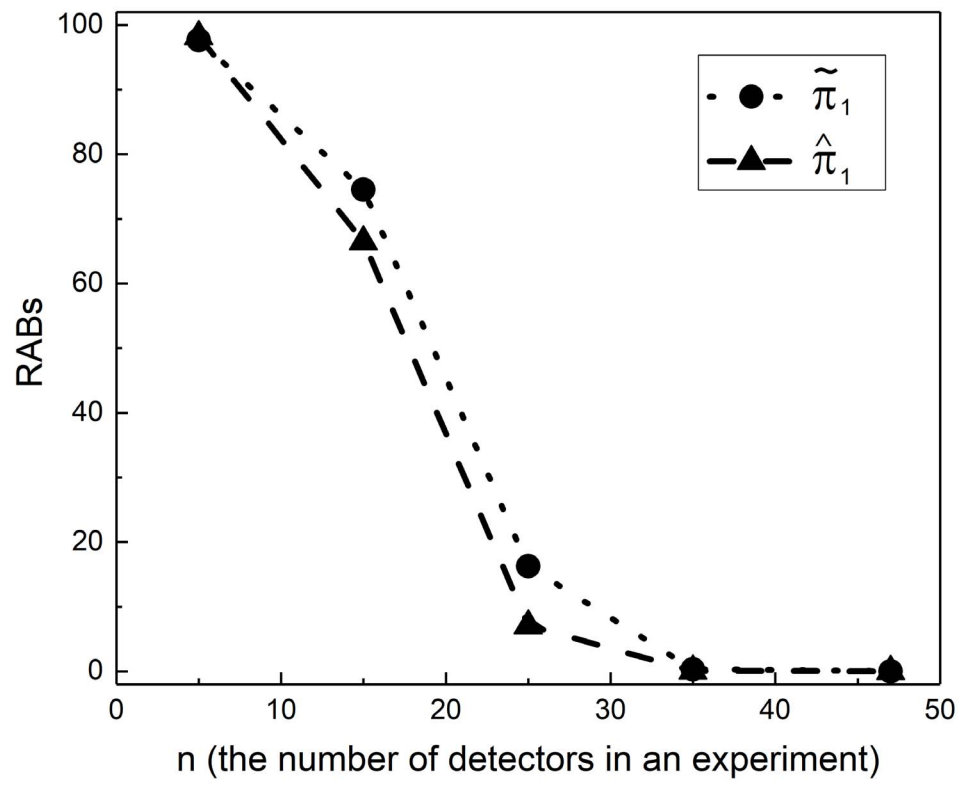

Fig. 3. Relative absolute bias (RABs, in %) of portion estimators of malicious files ($\pi_1 = 0.58579$) obtained in numerical experiments with truly heterogeneous $p_{+j}$'s and $p_{-j}$'s (Section III-E2).

In this setting, the values of each metric can vary over a wide range. This makes a direct comparison of their biases less revealing since the scale of the bias depends on the scale of the true metric value. Therefore, for ease of comparison we define the following *Relative Absolute Bias* (RAB) for the $j$-th estimator, $1 \leq j \leq n$:

$$\text{RAB}(j) = \frac{|\text{average bias of the estimator over all the samples}|}{\text{true value of the parameter being estimated}}, \quad (25)$$

which measures how accurate the estimator is relative to the true parameter value.

Table I summarizes the RABs of the other estimators for detectors $j = 3, 13, 23, 33, 43$ as examples. Fig. 3 summarizes the RABs of the $\pi_1$ estimators. It is worth mentioning that the RABs for the false positive probabilities ($p_{+j}$) of detectors 3 and 13 are high. This is because they have really small $p_{+j}$, namely $p_{+3} = 0.000617$ and $p_{+13} = 0.00938$. Although the scales of the biases are small, they are relatively large because the actual true values are very small. Overall, we can draw the following insight from Fig. 3 and Table I.

*Insight 3: In the case of* truly heterogeneous *$p_{+j}$'s and $p_{-j}$'s,*

- *majority voting can be ruined by the presence of a significant number of "poor" detectors.*
- *both the naïve estimators and the adjusted estimators can be accurate when the number of detectors is large. For example, when $n = 47$ detectors, the RABs of the two sets of estimators are the same, implying that majority voting achieves perfect accuracy and there is no room for further improvement over the adjusted estimators.*
- *when using $n < 47$ detectors, most RABs of the adjusted estimators are substantially lower than the RABs of the naïve estimators, implying that the adjusted estimators can indeed achieve a better accuracy.*

## IV. Application

### A. Dataset

Our dataset consists of 10,738,585 files collected from VirusTotal in 2015, involving 62 malware detectors in total. Since some files are not analyzed by every detector, we omit the detectors that labeled less than 100,000 files. Among the remaining detectors, TrendMicro and TrendMicro-HouseCall are apparently from the same vendor (i.e., TrendMicro) and McAfee and McAfee-GW-Edition are apparently from the same vendor (i.e., McAfee). Because two detectors from the same vendor would use some common technology (i.e., not independent of each other) and our statistical estimators assume that the detectors are independent of each other, we eliminate TrendMicro-HouseCall and McAfee-GW-Edition. Then, we further eliminate the files that are not labeled by all of the remaining detectors. As a result, we have 9,428,997 files that are labeled by 47 malware detectors, leading to a matrix $X_{ij}$ with 9,428,997 rows and 47 columns. These files correspond to 5 months in 2015: 3,760,291 (May), 2,430,201 (June), 344,067 (July), 1,918,299 (November), and 976,139 (December). This prompts us to analyze the entire dataset as a whole and analyze the corresponding 5 smaller datasets individually as well.

### B. Experimental Design

For the entire dataset and each of the 5 datasets, we consider the following sequence of experiments with different sets of detectors, which conduct the majority voting in each case.

- *Experiment 1*: We consider $n = 4$ handpicked detectors, namely Kaspersky, McAfee, Microsoft, and Symantec, because they are widely used in the real world.
- *Experiment 2*: In addition to the preceding 4 detectors, we randomly select 6 of the remaining detectors, leading to $n = 10$ detectors in the experiment.
- *Experiment 3*: In addition to the preceding 10 detectors, we randomly select 10 of the remaining detectors, leading to $n = 20$ detectors in the experiment.
- *Experiment 4*: In addition to the preceding 20 detectors, we randomly select 10 of the remaining detectors, leading to $n = 30$ detectors in the experiment.
- *Experiment 5*: In addition to the preceding 30 detectors, we randomly select 10 of the remaining detectors, leading to $n = 40$ detectors in the experiment.
- *Experiment 6*: We consider all of the $n = 47$ detectors.

### C. Estimating Metrics $p_{+j}, p_{-j}, q_{+j}, q_{-j}$

Fig. 4 plots the estimation of $p_{+j}, p_{-j}, q_{+j}, q_{-j}$ for each of the 47 detectors with respect to the entire (i.e., "All") and July datasets. Due to space limitations, we omit the May, June, and November datasets because they exhibit a phenomenon similar to the entire dataset as plotted in Fig. 4a, and we omit the December dataset because it exhibits a phenomenon similar to the July dataset as plotted in Fig. 4b.

TABLE I

RELATIVE ABSOLUTE BIAS (RABs, IN %) OF THE ESTIMATORS OF $p_{+j}, p_{-j}, q_{+j}, q_{-j}$ OBTAINED IN NUMERICAL EXPERIMENTS WITH TRULY HETEROGENEOUS $p_{+j}$'S AND $p_{-j}$'S (SECTION III-E2). NOTE THAT DETECTOR $j$ IS ENCOUNTERED ONLY WHEN $\kappa \geq j$ BECAUSE FOR AN EXPERIMENT WITH PARAMETER $\kappa$, ONLY THE FIRST $\kappa$ ($\kappa \leq n$) DETECTORS PARTICIPATE IN VOTING

| $\kappa$ | RAB($\tilde{p}_{+j}$) | RAB($\hat{p}_{+j}$) | RAB($\tilde{p}_{-j}$) | RAB($\hat{p}_{-j}$) | RAB($\tilde{q}_{+j}$) | RAB($\hat{q}_{+j}$) | RAB($\tilde{q}_{-j}$) | RAB($\hat{q}_{-j}$) |
|---|---|---|---|---|---|---|---|---|
| Detector $j = 3$ ($p_{+3} = 0.000617$, $p_{-3} = 0.828$, $q_{+3} = 0.99747$, $q_{-3} = 0.46047$) | | | | | | | | |
| 5 | 14535 | 15022 | 73.367 | 58.003 | 89.238 | 93.187 | 116.44 | 116.31 |
| 15 | 10814 | 10442 | 13.829 | 8.4348 | 57.131 | 52.283 | 91.456 | 81.263 |
| 25 | 2891.2 | 1700 | 2.0269 | 0.73005 | 8.9827 | 4.7222 | 20.906 | 8.9556 |
| 35 | 36.312 | 2.0026 | 0.1949 | 0.23764 | 0.097835 | 0.001765 | 0.31401 | 0.034455 |
| 47 | 10.004 | 10.004 | 0.14966 | 0.14966 | 0.023665 | 0.023665 | 0.077347 | 0.077347 |
| Detector $j = 13$ ($p_{+13} = 0.00938$, $p_{-13} = 0.339$, $q_{+13} = 0.99007$, $q_{-13} = 0.67387$) | | | | | | | | |
| 15 | 3277.7 | 3100.8 | 46.217 | 29.1 | 68.505 | 61.417 | 41.754 | 36.909 |
| 25 | 964.78 | 496.87 | 9.5812 | 2.94 | 12.133 | 5.5893 | 11.74 | 4.7638 |
| 35 | 8.877 | 3.6354 | 0.49269 | 0.19468 | 0.094029 | 0.035434 | 0.3934 | 0.13086 |
| 47 | 2.1952 | 2.1952 | 1.6503 | 1.6503 | 0.013516 | 0.013516 | 0.52851 | 0.52851 |
| Detector $j = 23$ ($p_{+23} = 0.0952$, $p_{-23} = 0.126$, $q_{+23} = 0.92849$, $q_{-23} = 0.83546$) | | | | | | | | |
| 25 | 136.57 | 74.2280 | 13.804 | 4.2669 | 14.676 | 6.981 | 5.4536 | 2.0588 |
| 35 | 3.0858 | 1.2075 | 0.27853 | 0.63963 | 0.282 | 0.10541 | 0.03217 | 0.097677 |
| 47 | 1.2355 | 1.2355 | 0.70389 | 0.70389 | 0.080952 | 0.080952 | 0.09464 | 0.09464 |
| Detector $j = 33$ ($p_{+33} = 0.19$, $p_{-33} = 0.044$, $q_{+33} = 0.87678$, $q_{-33} = 0.92866$) | | | | | | | | |
| 35 | 2.9361 | 1.2875 | 1.9487 | 2.6073 | 0.47275 | 0.1948 | 0.13101 | 0.19516 |
| 47 | 1.7409 | 1.7409 | 2.2299 | 2.2299 | 0.20125 | 0.20125 | 0.13066 | 0.13066 |
| Detector $j = 43$ ($p_{+43} = 0.227$, $p_{-43} = 0.0173$, $q_{+43} = 0.8596$, $q_{-43} = 0.96932$) | | | | | | | | |
| 47 | 0.3570 | 0.3570 | 3.8895 | 3.8895 | 0.059732 | 0.059732 | 0.12257 | 0.12257 |

We make the following observations. First, corresponding to the entire dataset, none of the detectors is *very good* (i.e., false positive and false negative probabilities are both smaller than 0.1). Therefore, 8 detectors whose false negative probabilities are greater than 0.5. More specifically, corresponding to the entire (thus, the May, June, and November) datasets, namely Fig. 4a, we observe that the false positive probabilities of the detectors, namely the red-colored $\tilde{p}_{+j}$'s and $\hat{p}_{+j}$'s, fall into the interval [0, 0.3]. Indeed, the adjusted minimum and maximum false positive probabilities are respectively 0.0006 (for a detector with a false negative probability of 0.8239) and 0.2756 (for a detector with a false negative probability of 0.0086). These two examples would manifest two different philosophies in designing detectors: trading a high false positive (negative) probability for a low false negative (positive) probability. Moreover, the Bayesian detection probabilities, namely the $\tilde{q}_{+j}$'s and $\hat{q}_{+j}$'s, fall into a relative small interval of [0.8, 1]. For the July (thus, the December) datasets, a similar phenomenon is exhibited by the Bayesian detection probabilities, but not by the false positive probabilities. However, the detectors' false negative probabilities, namely the $\tilde{p}_{-j}$'s and $\hat{p}_{-j}$'s, and negative predictive probability, namely the $\tilde{q}_{-j}$'s and $\hat{q}_{-j}$'s, vary substantially.

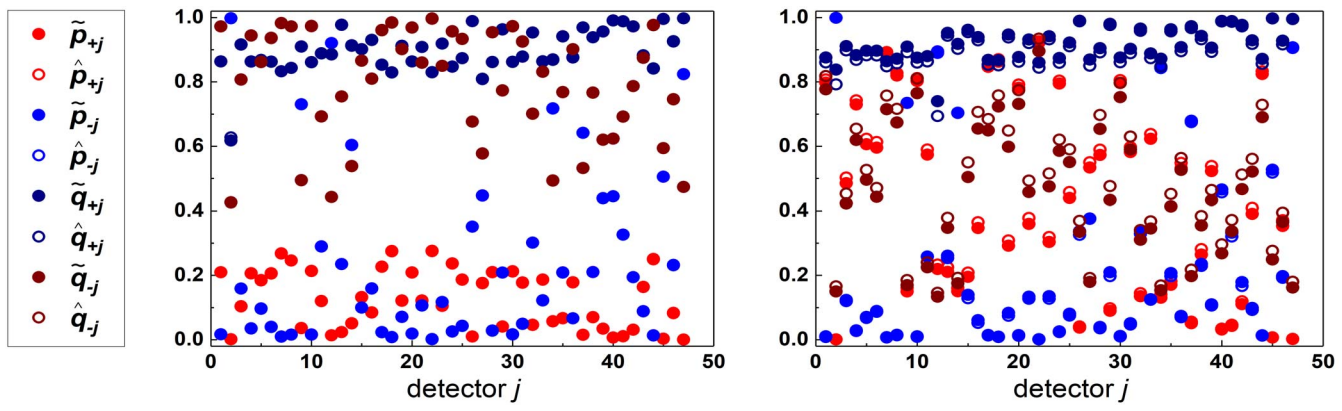

(a) (b)

Fig. 4. Estimation of $p_{+j}, p_{-j}, q_{+j}, q_{-j}$ for $j \in [1, 47]$ when using the 47 detectors to the entire dataset. (a) All (i.e., the entire dataset). (b) The July dataset.

Figs. 5 and 6 plot the estimation of $p_{+j}, p_{-j}, q_{+j}, q_{-j}$ for Kaspersky, McAfee, Microsoft, and Symantec, with respect to the entire and July datasets and with respect to the sequence of experiments of $n = 4, 10, 20, 30, 40, 47$ detectors mentioned above. We omit the May, June, and November datasets because they exhibit a phenomenon similar to the entire dataset in Fig. 5, and we omit the December dataset because it exhibits a phenomenon similar to the July dataset in Fig. 6. We make the following observations.

First, malware detectors achieve different trade-offs between their false positive probability and their false negative probability. Consider the case of $n = 47$ detectors with the entire (thus, the May, June and November) dataset as shown in Insight 3, we observe that the naïve estimators are almost the same as their adjusted estimators. This means that the estimators give accurate results. Kaspersky has a false positive probability $\tilde{p}_{+j} \approx \hat{p}_{+j} = 0.18632$, a false negative probability $\tilde{p}_{-j} \approx \hat{p}_{-j} = 0.04286$, a Bayesian detection probability $\tilde{q}_{+j} \approx \hat{q}_{+j} = 0.87382$ (i.e., when the Kaspersky detector says a file is malicious, the trustworthiness of the claim is only 87.382%), and a negative predictive probability $\tilde{q}_{-j} \approx \hat{q}_{-j} = 0.93370$ (i.e., when the Kaspersky detector says a file is benign, the trustworthiness of the claim is 93.370%). Notice that Kaspersky's false positive probability $\hat{p}_{+j}$ is almost 4 times of its false negative probability $\hat{p}_{-j}$. A similar phenomenon is exhibited by McAfee. On the other hand, Symantec has a smaller false positive probability and a larger false negative probability, namely $\tilde{p}_{+j} \approx \hat{p}_{+j} = 0.06956 < \tilde{p}_{-j} \approx \hat{p}_{-j} = 0.21011$. Moreover, Symantec has a larger Bayesian detection probability $\tilde{q}_{+j} \approx \hat{q}_{+j} = 0.93869$, and a smaller negative predictive probability, namely $\tilde{q}_{-j} \approx \hat{q}_{-j} = 0.76661$. These observations are consistent with $n = 10, 20, 30, 40, 47$ detectors, but not necessarily for $n = 4$ detectors, meaning that using the 4 popular detectors for majority voting can substantially and incorrectly underestimate the false positive probabilities of Kaspersky and McAfee. These observations invalidate the rule-of-thumb that anti-malware vendors often trade high false negative probabilities off for low false positive probabilities [8].

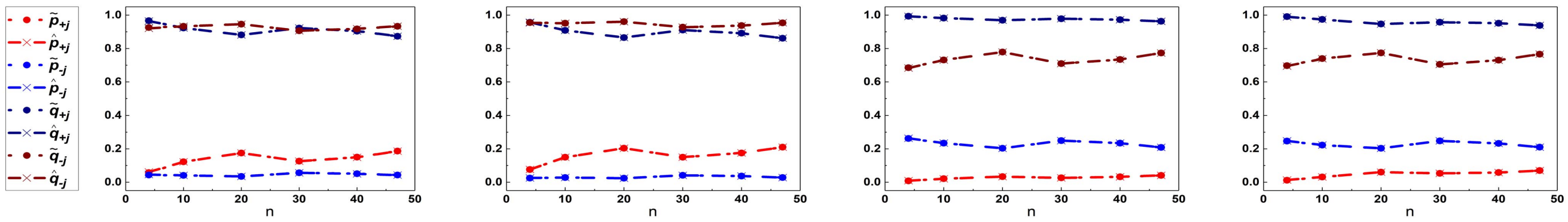

(a) (b) (c) (d)

Fig. 5. Estimation of $p_{+j}, p_{-j}, q_{+j}, q_{-j}$ for Kaspersky, McAfee, Microsoft, and Symantec when using $n = 4, 10, 20, 30, 40, 47$ detectors to the entire dataset. (a) Kaspersky. (b) McAfee. (c) Microsoft. (d) Symantec.

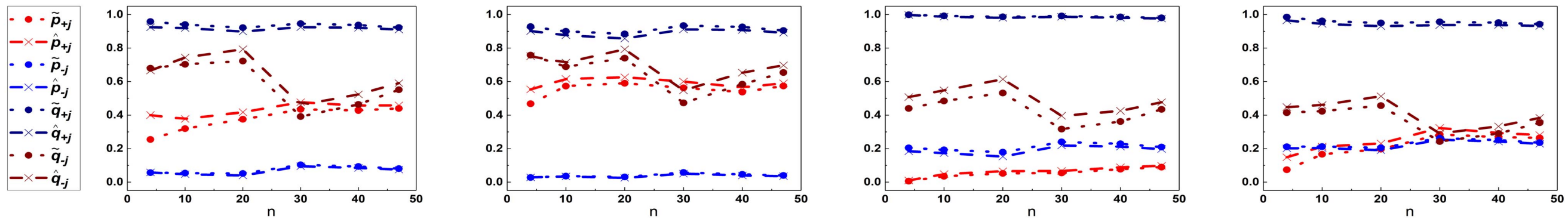

(a) (b) (c) (d)

Fig. 6. Estimation of $p_{+j}, p_{-j}, q_{+j}, q_{-j}$ for Kaspersky, McAfee, Microsoft, and Symantec when using $n = 4, 10, 20, 30, 40, 47$ detectors to the July dataset. (a) Kaspersky. (b) McAfee. (c) Microsoft. (d) Symantec.

Second, Fig. 6 shows that for the July (thus, the December) dataset, Kaspersky and McAfee also exhibit relatively low false negative probabilities, but very high false positive probabilities (e.g., $\hat{p}_{+j} = 0.4$ for Kaspersky and $\hat{p}_{+j} = 0.4$ for Kaspersky $\hat{p}_{+j} = 0.56$ for McAfee when $n = 47$). This justifies that during different periods of time, the detectors' detection capabilities can vary. When considering the entire dataset, the variation gets weighted down because the July and December datasets are much smaller than the datasets of the other months. Moreover, the adjusted estimators outperform their respective naïve estimators in most cases. This is because the naïve estimators have a large bias, as predicted by Theorem 2 that a large false positive probability $p_{+j}$ introduces a significant bias into the estimators.

### *D. Estimating the Fraction of Malicious Files $\pi_1$*

Fig. 7 plots $\tilde{\pi}_1$ and $\hat{\pi}_1$ (the $y$-axis) with respect to the number $n$ of detectors (the $x$-axis). We make the following observations. First, the fractions of malicious files in the 5 monthly datasets are different. Moreover, the $\tilde{\pi}_1$ of the entire dataset (i.e., the lines corresponding to "All" in Figure 7) is a weighted average of the $\tilde{\pi}_1$'s of the monthly datasets, based on the definition of $\tilde{\pi}_1$ in Eq. (8). Note that this linear relation does not hold for the adjusted estimators.

Second, for the entire dataset and the May, June, and November datasets, $\tilde{\pi}_1$ is almost identical to $\hat{\pi}_1$, meaning that they can be used as reliable estimators of the ground truth $\pi_1$. However, for the July and December datasets, there is a significant difference between $\hat{\pi}_1$ and $\tilde{\pi}_1$, meaning that $\hat{\pi}_1$ is a more accurate estimation of the ground truth $\pi_1$. Note that Eq. (19) indicates that the difference between $\hat{\pi}_1$ and $\tilde{\pi}_1$ is determined by $p_{01}$ and $p_{11}$. The discrepancy between $\hat{\pi}_1$ and $\tilde{\pi}_1$ for the July and December datasets is caused by, as shown in Section IV-C, the fact that $p_{01}$ is large and $p_{11} \approx 1$ for these two months. In contrast, for the entire dataset and the other three months, the discrepancy is small because $p_{01} \approx 0$ and $p_{11} \approx 1$. It is also interesting to note that despite the discrepancy between $\hat{\pi}_1$ and $\tilde{\pi}_1$ for the July and December datasets, the discrepancy for the entire dataset is almost negligible because the July and December datasets are substantially smaller than the three other months.

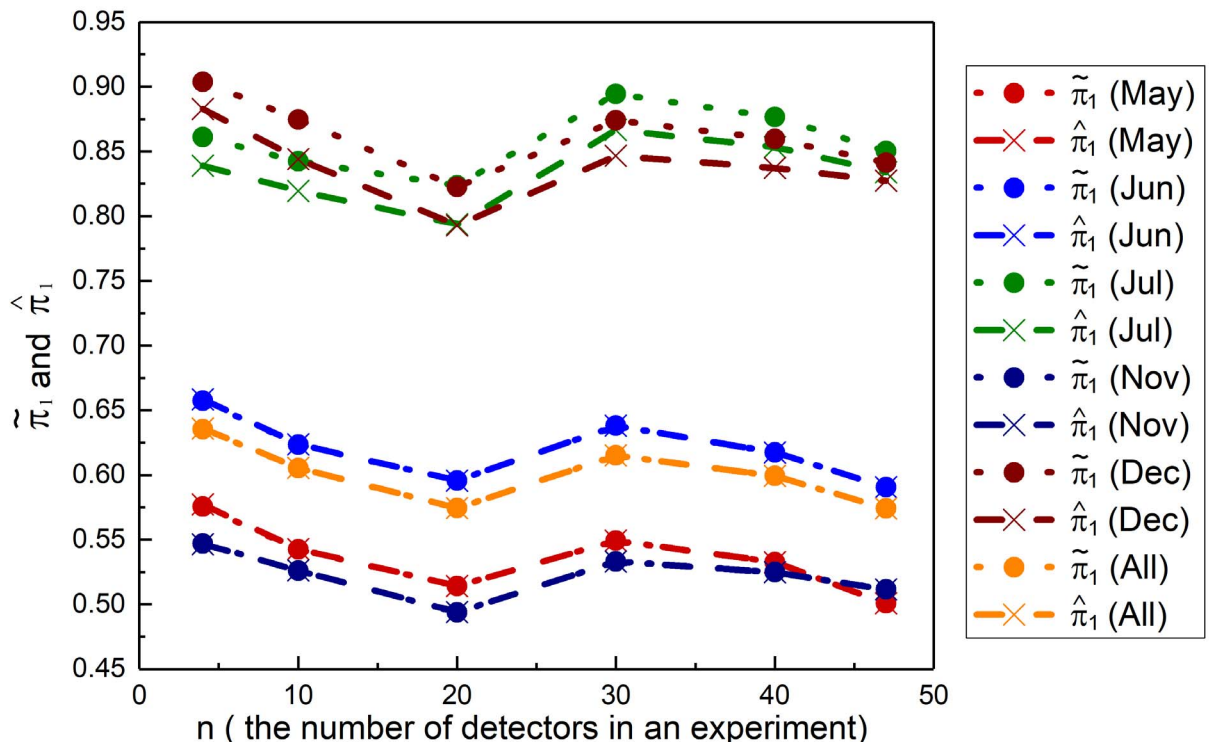

Fig. 7. Plots of $\tilde{\pi}_1$ (dotted lines) and $\hat{\pi}_1$ (dashed lines) with different datasets, where "all" means the entire dataset and two lines of the same color corresponds to the same dataset.

Third, Insight 3 implies that when the number of detectors is sufficiently large, the estimated value is almost the ground truth value. This means that for the entire dataset, we have $\pi_1 \approx \tilde{\pi}_1 = \hat{\pi}_1 = 0.58580$. Nevertheless, Fig. 7 exhibits a drop of both $\tilde{\pi}_1$ and $\hat{\pi}_1$ when $n = 20$. Our retrospective investigation shows that 5 of the 10 detectors that are newly added to Experiment 3 have medium-to-high false negatives, namely 0.998, 0.733, 0.447, 0.363, and 0.296 after rounding to 3 decimals. As a consequence, the voting result of these 20 detectors generates a lot more benign labels than the voting result of the 10 detectors in Experiment 2. Nevertheless, as more detectors are used, the converging trend of the estimators resumes. This further confirms the robustness of the majority voting method.

### E. Insights

First, we observe that $\pi_1 \approx \tilde{\pi}_1 = \hat{\pi}_1 = 0.57413$ when $n = 47$ detectors are used, and that the quite inaccurate $\hat{\pi}_1 = 0.63588$ is obtained when only using the 4 popular detectors (i.e., Kaspersky, McAfee, Microsoft, and Symantec). The same phenomenon is observed by the other estimators as well. For example, the false positive probability of Kaspersky in the case of using 47 detectors is $\hat{p}_{+j} = 0.18686$, which can be treated as the ground truth value; whereas, it is $\hat{p}_{+j} = 0.05763$ when only using the 4 popular detectors only. This leads to:

*Insight 4: The defender should use as many detectors as possible, rather than using a few popular detectors.*
This insight justifies the service paradigm of VirusTotal.

Second, once $\hat{p}_{+j}, \hat{p}_{-j}$, $j = 1, \ldots, n$, are available, the probability $p_{11}$ and the probability $p_{00}$ can be respectively estimated using the Monte Carlo method in Section III-D4 as

$$\hat{p}_{11} = \Pr\left(\sum_{j=1}^{n} X_{ij} \geq \frac{n}{2} | a_i = 1\right), \text{ where}$$
$$X_{ij} \sim Bernoulli(1 - \hat{p}_{-j}),$$
$$\hat{p}_{00} = \Pr\left(\sum_{j=1}^{n} X_{ij} < \frac{n}{2} | a_i = 0\right), \text{ where}$$
$$X_{ij} \sim Bernoulli(1 - \hat{p}_{+j}).$$

These two probabilities can also be combined to compute

$$\Pr(Y_i = a_i) = \hat{\pi}_1 \hat{p}_{11} + (1 - \hat{\pi}_1)\hat{p}_{00}, \tag{26}$$

which is the estimated average probability that the voted label of a file matches its true label. For example, in our application, when all the 47 detectors are used, we have $p_{11} = 1$ and $p_{00} = 1$ by using the Monte Carlo method with $N = 5,000,000$ (i.e., the voted label of a file is indeed its true label). This leads to:

*Insight 5: When the number of detectors are sufficiently large (e.g., $n = 47$), the estimated metrics can be used to further compute the probability that the voted label matches the corresponding true label according to Eq.* (26).

## V. Discussion

### A. Real-World Implications

The insights mentioned above highlight some real-world application scenarios. Insights 1-3 indicate (i) how many "very good" (i.e., $p_+ < 0.1$ and $p_- < 0.1$), "good" (i.e., $0.1 \leq p_+ < 0.2$ and $0.1 \leq p_- < 0.2$), or "fair" (i.e., $0.2 \leq p_+ < 0.5$ and $0.2 \leq p_- < 0.5$) detectors are needed in order to achieve almost perfect malware detection, and (ii) the use of "poor" (i.e., $p_+ \geq 0.5$ or $p_- \geq 0.5$) detectors can critically hurt the malware detection capability (i.e., such detectors should be avoided). For example, even if the detectors are "very good," at least 5 detectors are needed. However, we showed in Section IV that none of the 47 detectors is "very good." These insights also imply that the economic benefit of developing a smaller number of high quality detectors (i.e., "good" or "very good") may need to be reconsidered when the cost of lowering their false positive and false negative probabilities may not be linearly proportional to the gain in their detection capability. In summary, we have:

*Insight 6: If the detectors have (almost) identical $p_{+j}$ and (almost) identical $p_{-j}$, then 5 "very good" detectors can lead to (almost) perfect voting results. If the detectors have slightly different $p_+$'s and slightly different $p_{-j}$'s, 5 "very good" detectors and 15 "good" detectors can lead to (almost) perfect voting results. It is better to remove "poor" detectors in order to achieve more trustworthy results and then follow the guidelines in the identical and slightly heterogeneous cases to achieve trustworthy results.*

### B. Further Comparison With [8]

As mentioned in the Introduction, Kantchelian *et al.* [8] investigate both unsupervised and supervised learning approaches to the aggregation of the labels given by multiple malware detectors into a single one. The setting in their unsupervised learning approach is similar to the one considered in the present paper. However, they make 4 assumptions. In contrast, we only make 2 (of the 4) assumptions, which is made possible because we use a different approach. To be specific, [8] uses a *Bayesian approach*, in which all of the model parameters are considered as random quantities following certain distributions. In order to estimate a parameter (i.e., the mode of the parameter's distribution, namely the point having the highest probability), one needs to specify a prior distribution for the parameter, which essentially forces one to make further assumptions about the parameter. If the true parameter matches the assumed prior distribution, the Bayesian approach works well; otherwise, the Bayesian approach may fail to converge or it converges to some absurd estimator. Due to this reason, they have to assume the following prior distributions of the parameters: $\pi_1$ is around 0.5; and both true positive and false positive probabilities are low. In contrast, we take a *Frequentist approach*, in which the model parameters are not considered to be random. The estimator of a parameter is generally the optimizer of an objective function, representing the probability of seeing the data that has been observed. As such, no prior distributions on these parameters need to be assumed.

Now we present a numerical comparison between the estimators of [8] and ours. We apply their method to the synthetic data in the first experiment in Section III-E1 with $\pi_1 = 0.2$, $\epsilon = 0.1, 0.4$ and $n = 5, 15, 35$. That is, the false positive probabilities $p_{+j}$ and the false negative probabilities $p_{-j}$ of the $n$ detectors are randomly selected from the interval $[\epsilon, \epsilon + 0.1]$. Recall that a bias is defined as an average deviation of an estimator from the true value of the parameter in question, meaning that the bias can be positive or negative depending on which value is greater. For ease of comparison, we consider the absolute value of the bias, called absolute bias, which is denoted by |bias|. The closer the absolute bias of an estimator is to zero, the more accurate the estimator.

Fig. 8 plots the comparison between the absolute biases of their estimators and the absolute biases of our estimators,

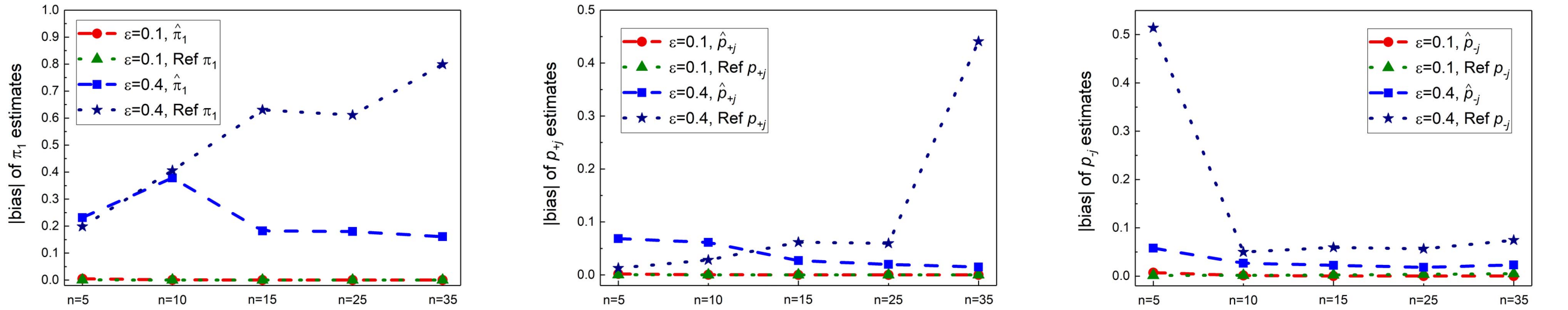

(a) (b) (c)

Fig. 8. Compare the absolute biases of the estimators of $\pi_1$, $p_{+j}$, $p_{-j}$ in [8] (indicated by "Ref" in the figures) and the adjusted estimators $\hat{\pi}_1$, $\hat{p}_{+j}$, $\hat{p}_{-j}$ in the present paper. (a) |Bias| of $\pi_1$. (b) |Bias| of $p_{+j}$. (c) |Bias| of $p_{-j}$.

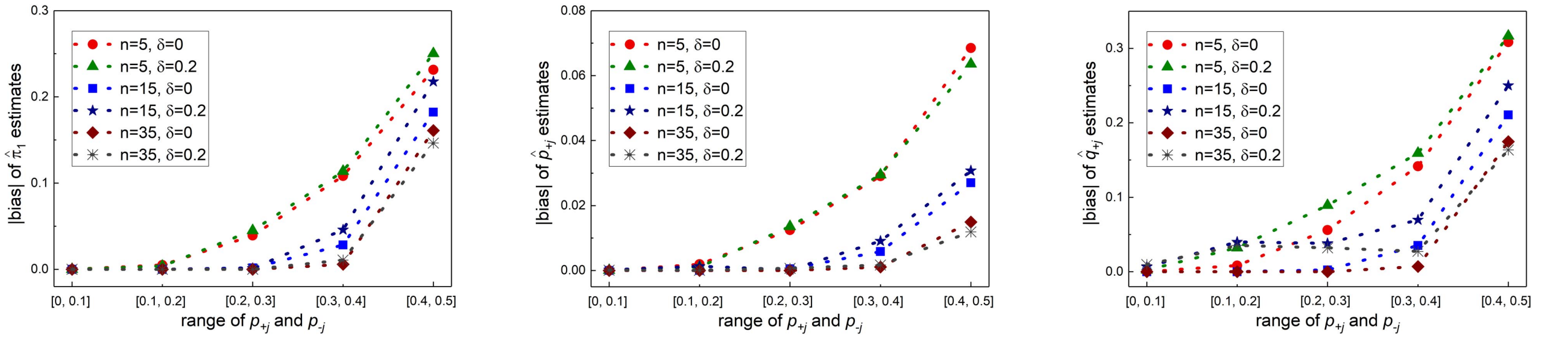

(a) (b) (c)

Fig. 9. Absolute biases of our adjusted estimators $\hat{\pi}_1$, $\hat{p}_{+j}$, $\hat{q}_{+j}$, where $\delta = 0$ corresponds to the data without "perturbing" the values of $p_{+j}$ and $p_{-j}$ over the individual files (i.e., Assumption 1 holds), and $\delta = 0.2$ corresponds to the data after "perturbing" the values of $p_{+j}$ and $p_{-j}$ with a range of $\pm 0.2$ (i.e., Assumption 1 is violated). (a) |Bias| of $\hat{\pi}_1$. (b) |Bias| of $\hat{p}_{+j}$. (c) |Bias| of $\hat{q}_{+j}$.

while noting that they do not have results on $q_{+j}$ and $q_{-j}$. We observe that when $\epsilon = 0.1$, all of the detectors have $p_{+j}$ and $p_{-j}$ between 0.1 and 0.2. We also observe that the absolute biases of both their estimators and ours are very close to 0 (i.e., $< 0.01$). Therefore, both their estimators and ours are accurate when $p_{+j}$ and $p_{-j}$ are sufficiently small (i.e., $0.1 \le p_{+j}, p_{-j} \le 0.2$).

When $\epsilon = 0.4$, all of the detectors are fair ones with $p_{+j}, p_{-j} \in (0.4, 0.5)$, which violates the low $p_{+j}$ assumption made in [8]. For every fixed $n$ except for $n = 10$, their method becomes erratic for estimating at least one of the three parameters. For example, when $n = 5$, their estimator for $p_{-j}$ has an absolute bias over 0.5; when $n = 15$ or 25, their estimator for $\pi_1$ has absolute biases around 0.6; when $n = 35$, their estimator for $\pi_1$ has an absolute bias of about 0.8, and their estimator for $p_{+j}$ has an absolute bias close to 0.5. The only stable case for their method is for $n = 10$, in which case the estimator for $p_{+j}$ has a slightly smaller absolute bias than that of our adjusted estimator; still, our estimators for $\pi_1$ and $p_{-j}$ have slightly smaller absolute biases than theirs. On the other hand, our estimators have consistently small biases for all the cases. In summary, our estimators have lower absolute biases than theirs in most cases. Even when our estimators have slightly larger absolute biases than theirs, the differences are all small.

We draw the following conclusion. The estimators in [8] are very sensitive to the assumption that all detectors have low false positive probabilities. When this assumption is violated, their estimators can become erratic and completely unreliable. On the other hand, our estimators are robust even when the detectors have high false positive probabilities.

### C. Are Our Estimators Sensitive to Assumption 1?

Assumption 1 is made in both [8] and the present paper. This assumption is needed for the theoretical derivation to be sound. The assumption says that a detector has the same false positive probability when classifying any benign file; and it has the same false negative probability when classifying any malicious file. This assumption can be relaxed by, as shown in Section IV, dividing the files according to their time-stamps, which is reasonable because the false positive and false negative probabilities indeed vary with the smaller datasets.

Now we explore whether Assumption 1 is absolutely necessary or not. We conduct experiments to see how sensitive our estimators are to the assumption. For this purpose, we first generate $p_{+j}$'s and $p_{-j}$'s according to the first experiment in Section III-E1 with $\pi_1 = 0.25$, $\epsilon = 0, 0.1, 0.2, 0.3, 0.4$ and $n = 5, 10, 15, 25, 35$. However, when generating the label of an individual file $i$ with respect to detector $j$, we use a false positive probability and a false negative probability randomly generated from the uniform distribution over $[p_{+j} - \delta, p_{+j} + \delta]$ and $[p_{-j} - \delta, p_{-j} + \delta]$, respectively. That is, we make the file-wise false positive and false negative probabilities vary in the range of $2\delta$. When the resulting probabilities are too close to 0 or 1, meaning that the probability interval would fall out of [0, 1], we truncate them to smaller intervals that are respectively centered at $p_{+j}$ and $p_{-j}$ while fitting into [0, 1]. We then apply our estimators to the resulting dataset. If the biases of the estimators corresponding to the perturbed data are not significantly different from the biases of the estimators corresponding to the unperturbed data, we can claim that the estimators are not sensitive to Assumption 1.

We conduct simulations with $\delta = 0.01, 0.02, 0.03, 0.04, 0.05, 0.1, 0.15, 0.2$. The larger the $\delta$, the bigger the deviation from the estimators of the unperturbed data. Therefore, we only report the results for $\delta = 0.2$ here in Fig. 9. Note that $\delta = 0$ corresponds to the case of unperturbed data. We observe that the largest difference between the biases corresponding to the perturbed data and the biases corresponding to the unperturbed data is observed for $\hat{\pi}_1$ and $\hat{q}_{+j}$ when $n = 15$ and the ranges of $p_{+j}$ and $p_{-j}$ are $[0.3, 0.4]$ or $[0.4, 0.5]$. When $p_{+j}$ and $p_{-j}$ are in these ranges, their perturbed versions at each file can take any values from a range of width 0.4 by our data generation method. However, even the largest difference between the biases is smaller than 0.05, while the true value is $\pi_1 = 0.2$ and the range of the true values of $q_{+j}$ is also between 0.2 and 1. This implies that the difference shows a high tolerance against the assumption, concluding that our estimators are not sensitive to Assumption 1.

### D. Limitations

Our study has the following limitations. First, we need to validate, weaken, or eliminate the assumptions. Assumption 1 says that an individual detector has the same capability in detecting all types of malware. However, some malware detectors may be particularly better at detecting some types of malware than others [5], [8]. Assumption 2 assumes that all detectors make their decisions independent of each other. This may not be universally true because Android malware detectors appear to have a weak correlation [15].

Second, we were able to quantify the bias of $\hat{\pi}_1$, but not able to quantify the bias of estimators $\hat{p}_{+j}$, $\hat{p}_{-j}$, $\hat{q}_{+j}$, and $\hat{q}_{-j}$ completely. Fully quantifying the biases of these estimators is a challenging task, but deserves a full investigation.

Third, we investigated the unweighted voting method. In practice, some detectors may be known to perform better than others. Therefore, one can investigate a weighted majority voting method.

## VI. Related Work

In addition to the most closely related prior work [8], we briefly review less related prior studies. Gu *et al.* [16] investigate alert fusion methods in the context of intrusion detection systems, called the fusion of decisions based on the alerts of multiple detectors, while assuming that the detectors are independent of each other. By considering the associated cost, they use the Neyman-Pearson detection theory and the Likelihood Ratio Test (LRT) [17] to identify the optimal fusion that incurs the minimum cost. However, they assume that the false positive probability $p_{+j}$ and false negative probability $p_{-j}$ of the $j$-th detector are given (e.g., computed from the known ground truth [16]). Their study and ours are complementary to each other in the following sense: Our work is to accurately estimate parameters (including $p_{+j}$ and $p_{-j}$) *without* knowing the ground truth; in contrast, their work assumes that $p_{+j}$ and $p_{-j}$ are given.

There are studies for dealing with inconsistent labels in the context of crowd-sourced labeling [10], [18]–[21]. However, these studies assume that some portions of the ground truth (e.g., the value of $a_i$ for some $i$'s, $p_{+j}$ for $j = 1, \ldots, n$, or $p_{-j}$ for $j = 1, \ldots, n$ when putting in the terminology of the present paper) are known. In contrast, these metrics are exactly what we aim to estimate.

Sebastian *et al.* [22] investigate how to automatically classify malware samples into different families. This problem is different from the one we study because of the following: Malware detectors can be seen as having two tasks: (i) decide whether a sample is a malware or not; and (ii) if the sample is a malware, to which malware family it belongs. We study problem (i), while *not* assuming the ground truth is known; they study problem (ii), while assuming that a sample that is detected as malware is indeed malware (i.e., the ground truth is known).

Hurier *et al.* [23] define a set of metrics to characterize the discrepancy between malware detectors while treating the *voting result* as the ground truth, namely the problem (i) mentioned above. In contrast, we differentiate the voting result from the ground truth, and aim to quantify the distance between the voting result and the unknown ground truth.

## VII. Conclusion

We motivated and formulated the measurement of malware detection metrics in the *absence* of ground truth as a *statistical estimation* problem. We presented a statistical methodology to tackle this problem by designing naïve estimators and adjusted estimators. We validated these estimators based on numerical experiments of synthetic data with known ground truth. We learned useful insights in terms of the accuracy (or usefulness) of these estimators in various parameter regimes. We applied these estimators to analyze a real dataset.

We also discussed the three limitations of the present study and correspondingly derived the following future research directions: (i) eliminating or weakening the assumptions; (ii) quantifying the bias of adjusted estimators; and (iii) developing an optimally weighted voting mechanism.

## Appendix A
## Proofs of the Theorems

*Theorem 1:* Note that $\sum_{i=1}^{m} Y_i = \sum_{i\in\mathcal{I}_1} Y_i + \sum_{i\in\mathcal{I}_0} Y_i$. The $Y_i$'s in $\sum_{i\in\mathcal{I}_1} Y_i$ are IID Bernoulli$(p_{11})$ random variables. Therefore, we have $\sum_{i\in\mathcal{I}_1} Y_i \sim$ Binomial$(m_1, p_{11})$. When $m_1 \to \infty$, this distribution can be approximated by Normal$(m_1 p_{11}, m_1 p_{11}(1 - p_{11}))$. Similarly, $\sum_{i\in\mathcal{I}_0} Y_i \sim$ Binomial$(m_0, p_{01})$, which can be approximated by Normal$(m_0 p_{01}, m_0 p_{01}(1 - p_{01}))$.

Note that $\sum_{i\in\mathcal{I}_1} Y_i$ and $\sum_{i\in\mathcal{I}_0} Y_i$ are independent of each other by our Assumption 2. Hence their sum $\sum_{i=1}^{m} Y_i$ asymptotically follows the Normal distribution with mean $m_1 p_{11} + m_0 p_{01}$ and variance $m_1 p_{11}(1 - p_{11}) + m_0 p_{01}(1 - p_{01})$. Therefore, the asymptotic distribution for $\tilde{\pi}_1$ is

$$\mu_1 = \frac{m_1 p_{11} + m_0 p_{01}}{m} = \pi_1 p_{11} + (1 - \pi_1) p_{01},$$

$$\sigma_1^2 = \frac{m_1 p_{11}(1 - p_{11}) + m_0 p_{01}(1 - p_{01})}{m^2} = \frac{\pi_1 p_{11}(1 - p_{11}) + (1 - \pi_1) p_{01}(1 - p_{01})}{m}.$$

**Algorithm 2** Data Generation for Simulation of Detectors With Slightly Heterogeneous False Positive and Negative Probabilities

INPUT: the number of detectors $n$, the lower bound $\epsilon$ of the probability range, $m = 50,000$, $\pi_1 = 0.2$, $N = 1,000$.

OUTPUT: $N$ datasets, each containing the $m \times n$ matrix $X$ of detector assigned labels, the size-$n$ vector of voted labels $Y$, the true malicious portion $\pi_1$ and the true metrics $p_{+j}, p_{-j}, q_{+j}, q_{+j}$, $j = 1, \ldots, n$ with $p_{+j}, p_{-j} \in (\epsilon, \epsilon + 0.1)$.

$m \leftarrow 50,000$
$\pi_1 \leftarrow 0.2$
$m_1 \leftarrow m\pi_1$
**for** $j = 1$ to $n$ **do**
  generate $p_{+j}$ and $p_{-j}$ such that $p_{+j} \sim \text{Uniform}(\epsilon, \epsilon + 0.1)$ and $p_{-j} \sim \text{Uniform}(\epsilon, \epsilon + 0.1)$
  $q_{+j} \leftarrow \pi_1(1 - p_{-j})/(\pi_1(1 - p_{-j}) + (1 - \pi_1)p_{+j})$
  $q_{-j} \leftarrow (1 - \pi_1)(1 - p_{+j})/((1 - \pi_1)(1 - p_{+j}) + \pi_1 p_{-j})$
**end for**
**for** $\ell = 1$ to $N$ **do**
  sample $m_1$ indices from $\{1, 2, \ldots, m\}$ to form the index set $\mathcal{I}_1$.
  **for** $i = 1$ to $m$ **do**
    **if** $i \in \mathcal{I}_1$ **then**
      $a_i \leftarrow 1$.
    **else**
      $a_i \leftarrow 0$.
    **end if**
  **end for**
  **for** $j = 1$ to $n$ **do**
    **for** $i = 1$ to $m$ **do**
      **if** $a_i = 1$ **then**
        generate $X_{ij}$ such that $X_{ij} \sim \text{Bernoulli}(1 - p_{-j})$.
      **else**
        generate $X_{ij}$ such that $X_{ij} \sim \text{Bernoulli}(p_{+j})$.
      **end if**
    **end for**
  **end for**
  **for** $i = 1$ to $m$ **do**
    **if** $\sum_{j=1}^{n} X_{ij} \geq \frac{n}{2}$ **then**
      $Y_i \leftarrow 1$.
    **else**
      $Y_i \leftarrow 0$.
    **end if**
  **end for**
**end for**
Output $N$ data sets as the final result.

This completes the proof.

*Theorem 2:* We prove the result for estimator $\tilde{p}_{+j}$. By the multivariate normal approximation to a multinomial distribution, we have

$$\left(\sum_{i\in\mathcal{I}_1} Z_{ij}^{(01)}, \sum_{i\in\mathcal{I}_1} Z_{ij}^{(00)}, \sum_{i\in\mathcal{I}_1} Z_{ij}^{(10)}, \sum_{i\in\mathcal{I}_1} Z_{ij}^{(11)}\right) \overset{\text{approx}}{\sim} N(\boldsymbol{\mu}_{1j}, \Sigma_{1j}),$$

$$\left(\sum_{i\in\mathcal{I}_0} Z_{ij}^{(01)}, \sum_{i\in\mathcal{I}_0} Z_{ij}^{(00)}, \sum_{i\in\mathcal{I}_0} Z_{ij}^{(10)}, \sum_{i\in\mathcal{I}_0} Z_{ij}^{(11)}\right) \overset{\text{approx}}{\sim} N(\boldsymbol{\mu}_{0j}, \Sigma_{0j}),$$

where for $c \in \{0, 1\}$, $\boldsymbol{\mu}_{cj} = (\mu_{c,01,j}, \mu_{c,00,j}, \mu_{c,10,j}, \mu_{c,11,j})$ and

$$\Sigma_{cj} = \begin{pmatrix} \sigma_{c,01,j}^2 & \rho_{c,01,00,j} & \rho_{c,01,10,j} & \rho_{c,01,11,j} \\ \rho_{c,01,00,j} & \sigma_{c,00,j}^2 & \rho_{c,00,10,j} & \rho_{c,00,11,j} \\ \rho_{c,01,10,j} & \rho_{c,00,10,j} & \sigma_{c,10,j}^2 & \rho_{c,10,11,j} \\ \rho_{c,01,11,j} & \rho_{c,00,11,j} & \rho_{c,10,11,j} & \sigma_{c,11,j}^2 \end{pmatrix}.$$

Let

$$\mathbf{Z} = \left(\sum_{i\in\mathcal{I}_1} Z_{ij}^{(01)}, \sum_{i\in\mathcal{I}_1} Z_{ij}^{(00)}, \sum_{i\in\mathcal{I}_1} Z_{ij}^{(10)}, \sum_{i\in\mathcal{I}_1} Z_{ij}^{(11)}, \sum_{i\in\mathcal{I}_0} Z_{ij}^{(01)}, \sum_{i\in\mathcal{I}_0} Z_{ij}^{(00)}, \sum_{i\in\mathcal{I}_0} Z_{ij}^{(10)}, \sum_{i\in\mathcal{I}_0} Z_{ij}^{(11)}\right).$$

According to Assumption 2, the two random vector components of $\mathbf{Z}$ are independent of each other. Hence we have

$$\mathbf{Z} \overset{\text{approx}}{\sim} N\left((\boldsymbol{\mu}_{1j}, \boldsymbol{\mu}_{0j}), \begin{pmatrix} \Sigma_{1j} & 0 \\ 0 & \Sigma_{0j} \end{pmatrix}\right).$$

For a vector $\mathbf{x} = (x_1, x_2, \ldots, x_8)$, consider function $g(\mathbf{x}) = (x_1 + x_5)/(x_1 + x_2 + x_5 + x_6)$. Then $\tilde{p}_{+j} = g(\mathbf{Z})$. Therefore, using the multivariate delta method to $\tilde{p}_{+j}$ yields the asymptotic distribution in the first part of the theorem. This completes the proof of the result for estimator $\tilde{p}_{+j}$. Proofs for the other estimators are similar and thus omitted.

## Appendix B
## Algorithm for Generating the Synthetic Data

See Algorithm 2.

## Acknowledgments

The authors would like to thank the reviewers for their constructive comments that guided the authors in improving the paper, VirusTotal for providing them the dataset, Moustafa Saleh and John Charlton for pre-processing the dataset, and Lisa Ho and John Charlton for proofreading the paper.

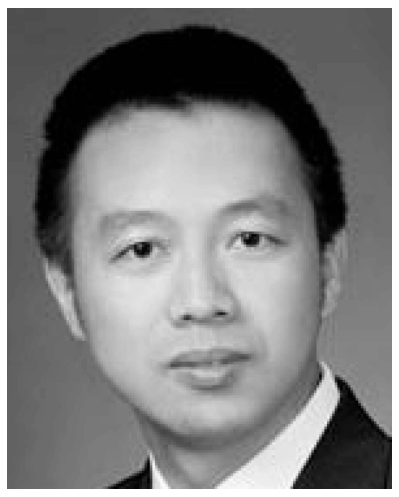

**Pang Du** received the B.S. and M.S. degrees in mathematics from the University of Science and Technology of China in 1996 and 1999, respectively, the M.A. degree in mathematics and the M.S.E. degree in computer science from Johns Hopkins University in 2002, and the Ph.D. degree in statistics from Purdue University in 2006. He joined the Department of Statistics, Virginia Tech, as an Assistant Professor, in 2006, and was promoted to Associate Professor with tenure in 2012, which is the position he currently holds. His current research interests include cybersecurity data analytics, analysis of functional and complex data, analysis of survival data and non-detects data, and statistical learning in high dimensional data.

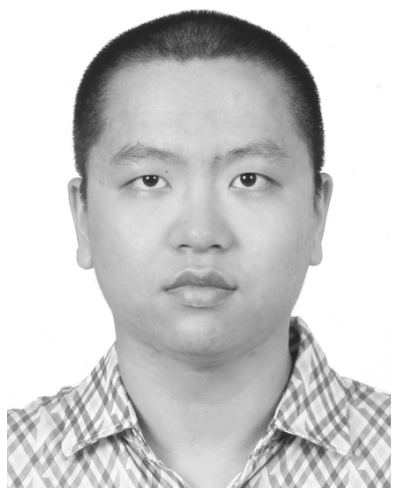

**Zheyuan Sun** received the M.S. degree in applied mathematics from the Illinois Institute of Technology. He is currently pursuing the Ph.D. degree with the Department of Computer Science, University of Texas at San Antonio. His research interests include cybersecurity data analytics.

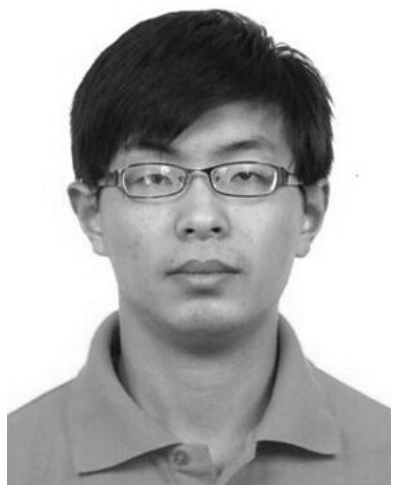

**Huashan Chen** received the M.S. degree from the Institute of Information Engineering, Chinese Academy of Sciences, in 2016. He is currently pursuing the Ph.D. degree with the Department of Computer Science, University of Texas at San Antonio. His primary research interests are in cybersecurity, especially moving target defense and security metrics.

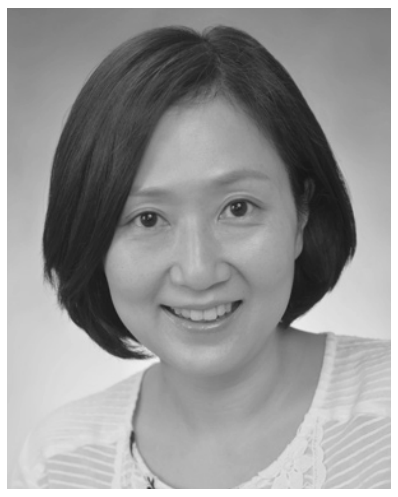

**Jin-Hee Cho** (SM’14) received the M.S. and Ph.D. degrees in computer science from Virginia Tech in 2004 and 2008, respectively. She is currently a Computer Scientist with the U.S. Army Research Laboratory, Adelphi, MD, USA. She has authored or co-authored over 90 peer-reviewed technical papers in leading journals and conferences in the areas of trust management, cybersecurity, network performance analysis, resource allocation, uncertainty reasoning and analysis, information fusion/credibility, and social network analysis. She received the Best Paper Award at IEEE TrustCom’2009, BRIMS’2013, IEEE GLOBECOM’2017, and 2017 ARL’s publication award. She was a recipient of the 2015 IEEE Communications Society William R. Bennett Prize in the field of communications networking. In 2016, she was selected for the 2013 Presidential Early Career Award for Scientists and Engineers, which is the highest honor bestowed by the U.S. Government on outstanding scientists and engineers in the early stages of their independent research careers.

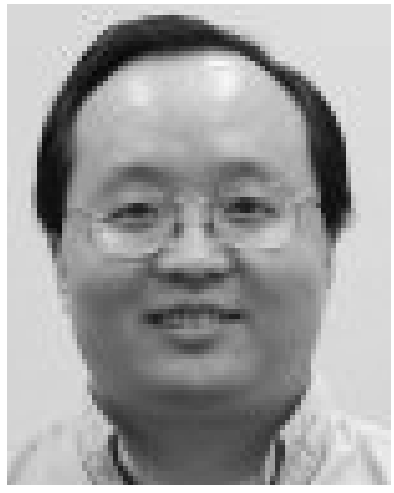

**Shouhuai Xu** received the Ph.D. degree in computer science from Fudan University. He is the Founding Director of the Laboratory for Cybersecurity Dynamics. He is also a Full Professor with the Department of Computer Science, University of Texas at San Antonio. He pioneered the Cybersecurity Dynamics Framework for modeling and analyzing cybersecurity from a holistic perspective (http://www.cs.utsa.edu/ shxu/socs). He co-initiated the International Conference on Science of Cyber Security (SciSec) in 2018 and the ACM Scalable Trusted Computing Workshop. He is/was the Program Committee Co-Chair of SciSec’18, ICICS’18, NSS’15, and Inscrypt’13. He was/is an Associate Editor of the IEEE TDSC, the IEEE T-IFS, and the IEEE TNSE.